%% file: main_v2_sjshin.tex
\documentclass[11pt]{article}
\input{head.tex}

\title{Sparse Sufficient Dimension Reduction for Block-Missing Multimodal Data}

\author{
Kim Jiyeop and Seung Jun Shin \\
\normalsize{Department of Statistics, Korea University}
}

\date{}

\begin{document}

\maketitle

\begin{abstract}
Integrating multiple data modalities increases predictor dimensionality, highlighting the need for dimension reduction. However, blockwise missingness complicates the application of many dimension reduction methods formulated as generalized eigenvalue problems. We propose a sparse dimension reduction framework for block-missing multimodal data. Our framework uses separate covariance shrinkage to construct method-specific matrices for the corresponding generalized eigenvalue problems, which are subsequently reformulated as penalized regression problems to induce sparsity. We further propose a prediction-based criterion for jointly selecting the covariance shrinkage and solver penalty parameters. Simulation studies across a range of dimension reduction methods, covariance structures, and missingness patterns demonstrate improved performance in both dimension reduction and variable selection. Finally, we apply the proposed framework to data from the Alzheimer's Disease Neuroimaging Initiative (ADNI) to illustrate its practical utility.
\end{abstract}

\section{Introduction}

Multimodal data consist of observations collected across multiple modalities and have become increasingly common across a variety of applications. For example, in healthcare applications, a single subject may be associated with demographic information, clinical measurements, and imaging data such as MRI and PET scans. Although different modalities can provide complementary information, combining them can substantially increase the predictor dimension, degrading the prediction performance of the subsequent models. This motivates the use of effective dimension reduction techniques for multimodal data.

Many dimension reduction methods can be expressed as a generalized eigenvalue problem involving the predictor covariance matrix. Their performance therefore depends strongly on reliable estimation of the covariance matrix. However, block-missingness is common in multimodal data as some modalities are expensive to acquire, subjects drop out during the study, or various datasets are combined. This makes the covariance estimation particularly challenging for multimodal data.

A naive approach to address block-missingness is to use only subjects with all modalities observed. However, this complete-case approach may substantially reduce the effective sample size and thereby undermine the benefits of integrating multiple modalities. Another natural option is to impute the missing values, but under block-missingness, the variables directly related to a missing value are often missing together as part of the same modality. Imputation must therefore rely more heavily on cross-modality associations or other structural assumptions, such as low-rank structure, and may perform poorly when these associations are weak or the assumptions are violated.

Available-case estimation provides an alternative by using all observations available for each moment. However, the resulting covariance matrix need not be positive definite, and cross-modality components may be unstable as they rely on fewer observations than within-modality components. To address these limitations, DISCOM \citep{yu2020discom} applies separate shrinkage to the within- and cross-modality covariance components and uses the resulting estimate in sparse linear regression. Related work by \citet{xiao2022block} incorporates the same shrinkage covariance estimator for sparse sliced inverse regression (SIR), using an adjusted linearized ADMM algorithm. 

In this work, we develop a unified framework that accommodates blockwise missingness and facilitates sparse dimension reduction in multimodal data. The key idea is to employ separate covariance shrinkage to estimate the matrices associated with a broad class of dimension reduction methods formulated as generalized eigenvalue problems. A regression-based reformulation of these problems allows us to incorporate sparsity-inducing penalties and perform dimension reduction and variable selection simultaneously. To determine the appropriate degree of regularization, we introduce a prediction-based criterion. Through simulations under various covariance structures and missingness patterns, we demonstrate the effectiveness of the proposed methods in terms of dimension reduction and variable selection. Its practical applicability is further illustrated through an analysis of the Alzheimer's Disease Neuroimaging Initiative (ADNI) data.

\section{Sparse Dimension Reduction for Block-Missing Data}
\label{sec:methodology}

\subsection{Dimension Reduction as a Generalized Eigenvalue Problem}
\label{sec:spectral-formulation}

Let $\mathbf{X}\in\mathbb{R}^p$ denote a predictor vector with mean $\boldsymbol{\mu}=E(\mathbf{X})$ and covariance matrix $\boldsymbol{\Sigma}=\operatorname{Cov}(\mathbf{X})$, and let $Y\in\mathbb{R}$ denote the response variable. Principal component analysis (PCA) \citep{pearson1901lines}, a widely used unsupervised dimension reduction method, seeks a low-dimensional representation that preserves as much predictor variation as possible. In contrast to PCA, sufficient dimension reduction (SDR) incorporates response information and seeks a full column rank matrix $\mathbf{B}\in\mathbb{R}^{p\times d}$, with $d<p$, such that $Y\perp\mathbf{X}\mid\mathbf{B}^\top\mathbf{X}$. Under this condition, $\mathbf{B}^\top\mathbf{X}$ retains all the information in $\mathbf{X}$ about $Y$, hence the term sufficient dimension reduction. Although the matrix $\mathbf{B}$ satisfying this conditional independence condition is not unique, under mild conditions, there exists a unique smallest subspace satisfying the condition. This subspace is called the central subspace and is the primary target of SDR.

Because directly estimating the full conditional distribution of $Y$ given $\mathbf{X}$ is difficult without strong structural assumptions, moment-based SDR methods instead construct estimators from inverse conditional moments, as in sliced inverse regression \citep[SIR,][]{li1991sir} and sliced averaged variance estimation \citep[SAVE,][]{cook1991save}, or principal Hessian direction \citep[PHD,][]{li1992phd}. Despite differences in their assumptions and specific targets, these methods admit a common generalized eigenvalue formulation \citep{li2007sparse}:
\begin{equation}
\label{eq:g_eign}
\mathbf{M}\boldsymbol{\phi}_j=\rho_j \mathbf{G}\boldsymbol{\phi}_j,\qquad j=1,\ldots,d,
\end{equation}
where $\rho_j$ and $\boldsymbol{\phi}_j$ are the $j$th eigenvalue and eigenvector. Here $\mathbf{M}\in\mathbb{R}^{p\times p}$ is a method-specific positive-semidefinite kernel matrix and $\mathbf{G}\in\mathbb{R}^{p\times p}$ is positive definite, often taking the form of the predictor covariance matrix. Assuming the eigenvalues are ordered, the leading $d$ eigenvectors span the resulting dimension reduction subspace, with $\boldsymbol{\phi}_i^\top\mathbf{G}\boldsymbol{\phi}_j=\delta_{ij}$, where $\delta_{ij}=1$ if $i=j$ and $0$ otherwise. Note that PCA can also be expressed as \eqref{eq:g_eign} with $(\mathbf{M},\mathbf{G})=(\boldsymbol{\Sigma},\mathbf{I}_p)$, where $\mathbf{I}_p$ denotes the $p$-dimensional identity matrix. Table~\ref{tab:generalized-eigen} summarizes the matrix pairs for various dimension reduction methods. Here, $\mathbf{Z}=\boldsymbol{\Sigma}^{-1/2}(\mathbf{X}-\boldsymbol{\mu})$ is the standardized predictor provided $\boldsymbol{\Sigma}\succ0$. We define $\boldsymbol{\Sigma}_{YZZ}=E[\{Y-E(Y)\}\mathbf{Z}\mathbf{Z}^\top]$ and $\boldsymbol{\Sigma}_{rZZ}=E(r\mathbf{Z}\mathbf{Z}^\top)$, where $r=Y-E(Y)-\{E(Y\mathbf{Z})\}^\top \mathbf{Z}$ is residual in the population linear regression.

\input{tb_dr_list}

\subsection{Sparse Generalized Eigenvalue Solver via Penalized Regression}
\label{sec:sparse-solver}

The eigenvectors $\boldsymbol{\phi}_j$, $j=1,\ldots,d$, in \eqref{eq:g_eign} are dense, making it difficult to interpret the estimated subspace and identify a set of relevant predictors. To impose sparcity, \citet{li2007sparse} proposed the following penalized regression problem for \eqref{eq:g_eign}:
\begin{equation}
\label{eq:opt_pen}
\begin{aligned}
(\boldsymbol{\alpha}^\star,\boldsymbol{\beta}^\star)
\in \arg\min_{\boldsymbol{\alpha},\boldsymbol{\beta}}\quad&
\sum_{i=1}^{p}
\left\|\mathbf{G}^{-1}\mathbf{m}_i-\boldsymbol{\alpha}\boldsymbol{\beta}^\top \mathbf{m}_i\right\|_{\mathbf{G}}^2
+\lambda_2\operatorname{tr}(\boldsymbol{\beta}^\top \mathbf{G}\boldsymbol{\beta})
+\sum_{j=1}^{d}\lambda_{1j}\|\boldsymbol{\beta}_j\|_1,\\
\text{subject to}\quad&
\boldsymbol{\alpha}^\top \mathbf{G}\boldsymbol{\alpha}=\mathbf{I}_d,
\end{aligned}
\end{equation},
where $\mathbf{m}_i$ is the $i$th column of the symmetric square root $\mathbf{M}^{1/2}$, $\boldsymbol{\alpha}_j$ and $\boldsymbol{\beta}_j$ are the $j$th columns of $\boldsymbol{\alpha}\in\mathbb{R}^{p\times d}$ and $\boldsymbol{\beta}\in\mathbb{R}^{p\times d}$, respectively. We write $\|\mathbf{u}\|_{\mathbf{G}}^2=\mathbf{u}^\top \mathbf{G}\mathbf{u}$. Without the last $\ell_1$-penalty term, the minimizer $\boldsymbol{\beta}_j^\star$ is proportional to $\boldsymbol{\phi}_j$, $j=1,\ldots,d$, provided $\rho_d>0$ \citep{li2007sparse}. The $\ell_1$-penalty induces sparsity for $\bbeta_j^\star$ and thus for $\bphi_j$.

We solve \eqref{eq:opt_pen} by alternately updating $\boldsymbol{\beta}$ and $\boldsymbol{\alpha}$. We initialize $\boldsymbol{\alpha}^{(0)}$ using a $\mathbf{G}$-orthonormal basis for the subspace estimated by the corresponding nonsparse method. At iteration $t$, with $\boldsymbol{\alpha}^{(t)}$ held fixed, each vector $\boldsymbol{\beta}_j$ is updated independently by solving
\[
\boldsymbol{\beta}_j^{(t+1)}
=\arg\min_{\boldsymbol{\beta}_j\in\mathbb{R}^p}
\left\{
\boldsymbol{\beta}_j^\top(\mathbf{M}+\lambda_2\mathbf{G})\boldsymbol{\beta}_j
-2(\boldsymbol{\alpha}_j^{(t)})^\top \mathbf{M}\boldsymbol{\beta}_j
+\lambda_{1j}\|\boldsymbol{\beta}_j\|_1
\right\}.
\]
Equivalently,
\[
\boldsymbol{\beta}_j^{(t+1)}
=\arg\min_{\boldsymbol{\beta}_j\in\mathbb{R}^p}
\left\{
\|\mathbf{u}_j^{*(t)}-\mathbf{M}^*\boldsymbol{\beta}_j\|_2^2
+\lambda_{1j}\|\boldsymbol{\beta}_j\|_1
\right\},
\]
where
\[
\mathbf{M}^*=
\begin{bmatrix}
\mathbf{M}^{1/2}\\
\sqrt{\lambda_2}\mathbf{G}^{1/2}
\end{bmatrix},
\qquad
\mathbf{u}_j^{*(t)}=
\begin{bmatrix}
\mathbf{M}^{1/2}\boldsymbol{\alpha}_j^{(t)}\\
\mathbf{0}_p
\end{bmatrix}.
\]
This is a standard lasso problem and it can be solved efficiently by coordinate descent \citep{friedman2010regularization} or the LARS \citep{efron2004least} algorithm.

Next, with $\boldsymbol{\beta}^{(t+1)}$ being fixed, we update $\boldsymbol{\alpha}$ by solving a weighted orthogonal Procrustes problem. Given the thin singular value decomposition
\[
\mathbf{G}^{-1/2}\mathbf{M}\boldsymbol{\beta}^{(t+1)}
=\mathbf{U}\mathbf{S}\mathbf{V}^\top,
\]
we set
\[
\boldsymbol{\alpha}^{(t+1)}
=\mathbf{G}^{-1/2}\mathbf{U}\mathbf{V}^\top.
\]
These two steps are repeated until the objective function value stabilizes.

The final estimates for $\bB$ is given by the normalized $\hat \bbeta$ where $\widehat{\boldsymbol{\beta}}$ denotes the sample minimizer of \eqref{eq:opt_pen}, i.e., 
\[
\widehat{\mathbf{B}}=
\begin{bmatrix}
\widehat{\boldsymbol{\beta}}_1/\|\widehat{\boldsymbol{\beta}}_1\|_2
& \cdots &
\widehat{\boldsymbol{\beta}}_d/\|\widehat{\boldsymbol{\beta}}_d\|_2
\end{bmatrix}.
\]

\subsection{Construction of Method-Specific Matrices under Block Missingness}
\label{sec:matrix-construction}

Now, the challenge is to accurately estimate the method-specific matrices $\mathbf{M}$ and $\mathbf{G}$ in Table~\ref{tab:generalized-eigen} under block missingness. We first start with available-case moment estimators, then construct a shrinkage covariance estimator by applying separate shrinkage to the available-case covariance
matrix. These estimates are used to construct the eigenvalue problems for a variety of dimension reduction methods, such as SIR, SAVE, PHD as well as PCA. 

Let $x_{ij}$ denote the value of predictor $j$ for subject $i$. Define the observation set $\mathcal{S}_j$ as the set of subject indices for which predictor $j$ is observed. The joint observation set $\mathcal{S}_{jt}=\mathcal{S}_j\cap\mathcal{S}_t$ contains subjects for whom both predictors $j$ and $t$ are observed. Write $n_j=|\mathcal{S}_j|$ and $n_{jt}=|\mathcal{S}_{jt}|$ for the corresponding sample sizes. Under block missingness, predictors within the same modality share an observation set. We assume that the responses, $y_i$s are fully observed.

The available-case moment estimates for predictor means and covariances are
\[
\widetilde\mu_j^{(jt)}
=\frac{1}{n_{jt}}\sum_{i\in\mathcal{S}_{jt}}x_{ij},
\qquad
\widetilde\sigma_{jt}
=\frac{1}{n_{jt}-1}
\sum_{i\in\mathcal{S}_{jt}}
(x_{ij}-\widetilde\mu_j^{(jt)})
(x_{it}-\widetilde\mu_t^{(jt)}).
\]
Note that when $j=t$, $\mathcal{S}_{jj}=\mathcal{S}_j$ and $n_{jj}=n_j$; we write $\widetilde\mu_j=\widetilde\mu_j^{(jj)}$ for simplicity. Collecting these estimates gives
$\widetilde{\boldsymbol{\mu}} = [\widetilde\mu_1,\ldots,\widetilde\mu_p]^\top$
and
$\widetilde{\boldsymbol{\Sigma}} =[\widetilde\sigma_{jt}]_{j,t=1}^{p}$.

Available-case estimation is appealing as it uses all available observations for each variable pair. However, it has two limitations: the resulting covariance matrix need not be positive definite, and cross-modality components may have greater estimation variability because they are based on substantially fewer observations than within-modality components. To address these limitations, \citet{yu2020discom} proposed a block-specific shrinkage covariance estimator. Partition $\widetilde{\boldsymbol{\Sigma}}$ according to the $K$ modalities and write
\[
\widetilde{\boldsymbol{\Sigma}}=
\begin{bmatrix}
\widetilde{\boldsymbol{\Sigma}}_{11} & \cdots & \widetilde{\boldsymbol{\Sigma}}_{1K}\\
\vdots & \ddots & \vdots\\
\widetilde{\boldsymbol{\Sigma}}_{K1} & \cdots & \widetilde{\boldsymbol{\Sigma}}_{KK}
\end{bmatrix},
\]
where $\widetilde{\boldsymbol{\Sigma}}_{kk}$ contains covariance estimates within modality $k$, whereas
$\widetilde{\boldsymbol{\Sigma}}_{k\ell}$ contains cross-modality covariance estimates between modalities $k$ and $\ell$ for $k\ne\ell$. Define
\[
\widetilde{\boldsymbol{\Sigma}}_I=
\begin{bmatrix}
\widetilde{\boldsymbol{\Sigma}}_{11} & \cdots & \mathbf{0}\\
\vdots & \ddots & \vdots\\
\mathbf{0} & \cdots & \widetilde{\boldsymbol{\Sigma}}_{KK}
\end{bmatrix},
\qquad
\widetilde{\boldsymbol{\Sigma}}_C
=\widetilde{\boldsymbol{\Sigma}}
-\widetilde{\boldsymbol{\Sigma}}_I.
\]
Thus, $\widetilde{\boldsymbol{\Sigma}}_I$ and $\widetilde{\boldsymbol{\Sigma}}_C$ contain the intra- and
cross-modality covariance components, respectively. The shrinkage estimator proposed by \citet{yu2020discom} is 
\begin{equation}
\label{eq:discom_est}
\widehat{\boldsymbol{\Sigma}}
=\alpha_1\widetilde{\boldsymbol{\Sigma}}_I
+\alpha_2\widetilde{\boldsymbol{\Sigma}}_C
+\alpha_3\mathbf{I}_p,
\qquad \alpha_1,\alpha_2, \alpha_3\in[0,1].
\end{equation}
The separate weights account for differences in estimation accuracy, while sufficient shrinkage toward the identity ensures positive definiteness. The selection of shrinkage parameters $\nu_1$, $\nu_2$ and $\nu_3$ are discussed in Section~\ref{sec:prediction-evaluation}.

We now construct method-specific kernel matrices using the shrinkage covariance estimator and available-case moments. First, we express the population kernels in Table~\ref{tab:generalized-eigen} in the original coordinates using $\mathbf{Z}=\boldsymbol{\Sigma}^{-1/2} (\mathbf{X}-\boldsymbol{\mu})$. This allows the required moments to be estimated without directly whitening incomplete observations. We then replace $\boldsymbol{\Sigma}$ by $\widehat{\boldsymbol{\Sigma}}$ defined in \eqref{eq:discom_est} and estimate the remaining moments using all observations available for each moment.

Table~\ref{tab:estimated-pairs} summarizes the resulting matrix estimates for the eigenvalue problem \eqref{eq:g_eign}.  Here, $\widehat p_h$ is the proportion of observations in response slice $h$, and $\widetilde{\boldsymbol{\mu}}_{(h)}$ and 
$\widetilde{\boldsymbol{\Gamma}}_{(h)}$ are the available-case mean vector and covariance matrix for that slice.
The matrices $\widetilde{\mathbf{T}}_Y$ and $\widetilde{\mathbf{T}}_r$ estimate the second moments of
$\mathbf{X}-\boldsymbol{\mu}$ weighted by the centered response and the population linear regression residual, respectively. For PCA, we take $\widehat{\mathbf{M}}=\widehat{\boldsymbol{\Sigma}}$ and $\widehat{\mathbf{G}}=\mathbf{I}_p$. Explicit formulas and derivations for the kernels are provided in Appendix~\ref{app:matrix-derivations}. When $\widehat{\boldsymbol{\Sigma}}\succ0$, all listed kernel matrices are positive semidefinite, and the resulting matrix pairs can be supplied directly to the sparse solver.

\input{tb_estimated_pairs}

Note that we use a plug-in kernel construction based on the shrinkage covariance estimator and available-case moments. This is reasonable because elementary available-case moment estimators can be unbiased and consistent under suitable conditions, while covariance shrinkage can improve numerical stability when inverting the covariance matrix.
Further kernel regularization may require method-specific choices
and additional tuning parameters, and we leave such refinements
for future work. Refined estimators can be incorporated
without changing the solver, provided that the resulting
$\widehat{\mathbf{M}}$ is positive semidefinite and
$\widehat{\mathbf{G}}$ is positive definite.

\subsection{Sparse Dimension Reduction Algorithm for Block-Missing Data}
\label{sec:fixed-fitting}

Algorithm~\ref{alg:fixed-fitting} describes the sparse dimension
reduction algorithm for block-missing data. Given the shrinkage
parameters $(\alpha_1,\alpha_2)$, with $\alpha_3$ determined by
\eqref{eq:practical-shrinkage}, and solver penalty parameters
$(\boldsymbol{\lambda}_1,\lambda_2)$, it combines method-specific
matrix construction with the sparse solver to estimate the
sparse dimension reduction subspace. The target dimension $d$ is specified. $D$ denotes the chosen method among SIR, SAVE, PHD, and PCA, and $\mathcal{D}$ denotes
the data used for fitting.

\algrenewcommand{\algorithmicrequire}{\textbf{Input:}}
\algrenewcommand{\algorithmicensure}{\textbf{Output:}}

\begin{algorithm}[H]
\caption{Sparse dimension reduction for block-missing data}
\label{alg:fixed-fitting}
\begin{algorithmic}[1]
\Require Data $\mathcal{D}$, method $D$, dimension $d$,
and admissible hyperparameters $(\alpha_1,\alpha_2,\boldsymbol{\lambda}_1,\lambda_2)$
\Ensure Sparse loading matrix $\widehat{\mathbf{B}}$
\State Estimate the available-case covariance matrix
$\widetilde{\boldsymbol{\Sigma}}$ from $\mathcal{D}$.
\State Construct $\widehat{\boldsymbol{\Sigma}}$ using
$(\alpha_1,\alpha_2)$ according to \eqref{eq:practical-shrinkage}.
\State Form the matrix pair
$(\widehat{\mathbf{M}},\widehat{\mathbf{G}})$
for method $D$ according to Table~\ref{tab:estimated-pairs}.
\State Apply the sparse solver in Section~\ref{sec:sparse-solver}
to $(\widehat{\mathbf{M}},\widehat{\mathbf{G}})$
with penalties $(\boldsymbol{\lambda}_1,\lambda_2)$ to obtain $d$ sparse loadings.
\State Normalize each loading vector to unit Euclidean norm
and collect them in $\widehat{\mathbf{B}}$.
\State \Return $\widehat{\mathbf{B}}$.
\end{algorithmic}
\end{algorithm}

\section{Hyperparameter Selection}
\label{sec:tuning}

Algorithm~\ref{alg:fixed-fitting} requires the shrinkage parameters
$(\alpha_1,\alpha_2)$ for matrix construction and the penalty
parameters $(\boldsymbol{\lambda}_1,\lambda_2)$ for the sparse
solver. In this section, we propose a criterion
based on prediction error in the estimated subspace to jointly
select the shrinkage and solver penalty parameters in practice.

\subsection{Tuning Criterion and Implementation}
\label{sec:prediction-evaluation}

The proposed tuning criterion follows an important motivation
for SIR \citep{li1991sir}: to make nonparametric regression
more feasible when the original predictor space suffers
from the curse of dimensionality. The central idea is to
identify a low-dimensional subspace that retains response
information, so that we can predict the response accurately
using nonparametric regression on the projected predictors.
Following this principle, we assess each estimated subspace
by how accurately the response can be predicted using the
projected predictors. 

For each estimated loading matrix $\widehat{\mathbf{B}}$,
we first form the projected predictors
$\widehat{\mathbf{B}}^\top\mathbf{x}_i$, $i=1,\ldots,n_v$,
where $\mathbf{x}_i$ denotes the predictor vector for
validation observation $i$. Let $\widehat f$ denote a fitted
nonparametric regression function of the response using these projected
predictors. We evaluate the estimated subspace by its mean squared
prediction error:
\begin{equation}
\label{eq:validation-risk}
\widehat{\mathcal{R}}_v
=\frac{1}{n_v}\sum_{i=1}^{n_v}
\left\{
y_i-\widehat f(\widehat{\mathbf{B}}^\top\mathbf{x}_i)
\right\}^{2},
\end{equation}
where $n_v$ is the validation sample size. For the prediction model, we use leave-one-out
$k$-nearest-neighbor (KNN) regression as a simple implementation
choice, with $k=\lceil\sqrt{n_v}\rceil$. Appendix~\ref{app:reduced-risk} shows that being a sufficient
subspace is sufficient, but not necessary, for attaining
the minimum prediction risk under squared-error loss
when each subspace is evaluated using its optimal population
predictor.

One consideration is that the same subspace can be
represented by different bases. These bases may yield different
Euclidean distances between projected observations and hence
different prediction errors from distance-based models,
such as KNN regression or Nadaraya--Watson regression with a
radial kernel. We therefore orthonormalize each loading matrix
before evaluation, making the criterion well-defined
independently of the choice of basis.
A proof is given in Appendix~\ref{app:basis-invariance}.
Orthonormalization is used only for evaluation; the sparse
loading matrix is returned.

We use a common sparsity parameter
$\lambda_{1j}=\lambda_1$, $j=1,\ldots,d$, following the
practical choice in \citet{li2007sparse}, and retain
$\lambda_1$ and $\lambda_2$ as solver penalty parameters.
\citet{yu2020discom} show that the oracle shrinkage weights under
Frobenius loss are
\begin{equation}
\label{eq:oracle-weights}
\alpha_1^\star
=\frac{\theta^2}{\theta^2+\delta_I^2},
\qquad
\alpha_2^\star
=\frac{\|\boldsymbol{\Sigma}_C\|_F^2}
{\|\boldsymbol{\Sigma}_C\|_F^2+\delta_C^2},
\qquad
\alpha_3^\star=(1-\alpha_1^\star)\gamma,
\end{equation}
where $\gamma=\operatorname{tr}(\boldsymbol{\Sigma})/p$,
$\theta^2=\|\boldsymbol{\Sigma}_I-\gamma\mathbf{I}_p\|_F^2$,
$\delta_I^2=E\|\widetilde{\boldsymbol{\Sigma}}_I
-\boldsymbol{\Sigma}_I\|_F^2$, and
$\delta_C^2=E\|\widetilde{\boldsymbol{\Sigma}}_C
-\boldsymbol{\Sigma}_C\|_F^2$.

Using the relationship for $\alpha_3^\star$, we set
$\alpha_3=(1-\alpha_1)\widehat\gamma$, as in DISCOM
\citep{yu2020discom}, with
$\widehat\gamma=
\operatorname{tr}(\widetilde{\boldsymbol{\Sigma}})/p$.
This plug-in parameterization gives
\begin{equation}
\label{eq:practical-shrinkage}
\widehat{\boldsymbol{\Sigma}}(\alpha_1,\alpha_2)
=
\alpha_1\widetilde{\boldsymbol{\Sigma}}_I
+\alpha_2\widetilde{\boldsymbol{\Sigma}}_C
+(1-\alpha_1)\widehat\gamma\mathbf{I}_p.
\end{equation}

We consider $(\alpha_1,\alpha_2)\in[0,1]^2$ and
$\lambda_1,\lambda_2>0$, retaining only configurations
satisfying
$\lambda_{\min}\{
\widehat{\boldsymbol{\Sigma}}(\alpha_1,\alpha_2)\}
\geq\varepsilon$
for a small positive threshold $\varepsilon$.
This condition ensures positive definiteness and is
checked again during final refitting.

PCA fits within Algorithm~\ref{alg:fixed-fitting},
but we do not apply the supervised tuning criterion
to it. Its shrinkage weights may be chosen heuristically
or through plug-in estimates of the oracle weights
in \eqref{eq:oracle-weights}, which we do not discuss here. The penalty parameters
can then be explored over a grid to obtain a desired
sparsity level.

\subsection{Complete Algorithm}
\label{sec:algorithm}

Algorithm~\ref{alg:block-missing-dr} summarizes the complete
sparse SDR procedure. For a given SDR method and each
hyperparameter configuration, it calls
Algorithm~\ref{alg:fixed-fitting} on the training data
$\mathcal{D}_{\mathrm{tr}}$ and evaluates the resulting
subspace on the same fully observed validation data
$\mathcal{D}_v$. After selecting the hyperparameters,
it refits on the combined data
$\mathcal{D}_{\mathrm{tr}}\cup\mathcal{D}_v$.

\begin{algorithm}[H]
\caption{Complete sparse SDR algorithm with hyperparameter selection}
\label{alg:block-missing-dr}
\begin{algorithmic}[1]
\Require Data $\mathcal{D}$, SDR method $D$,
dimension $d$, and hyperparameter grid $\Theta$
\Ensure Final sparse loading matrix $\widehat{\mathbf{B}}$

\State Reserve a fully observed validation subset
$\mathcal{D}_v$ and use the remaining data as
$\mathcal{D}_{\mathrm{tr}}$.

\For{each admissible hyperparameter configuration in $\Theta$}
    \State Apply Algorithm~\ref{alg:fixed-fitting} to
    $\mathcal{D}_{\mathrm{tr}}$ using the current
    configuration to obtain $\widehat{\mathbf{B}}$.
    \State Replace $\widehat{\mathbf{B}}$
    with an orthonormal basis for
    $\operatorname{col}(\widehat{\mathbf{B}})$
    obtained by QR decomposition.
    \State Compute and record the LOO-KNN mean squared prediction
    error in \eqref{eq:validation-risk} on $\mathcal{D}_v$.
\EndFor

\State Select the configuration with the lowest validation
error among those admissible on
$\mathcal{D}_{\mathrm{tr}}\cup\mathcal{D}_v$.
\State Apply Algorithm~\ref{alg:fixed-fitting} to
$\mathcal{D}_{\mathrm{tr}}\cup\mathcal{D}_v$
using the selected configuration to obtain the final
$\widehat{\mathbf{B}}$.
\State \Return $\widehat{\mathbf{B}}$.
\end{algorithmic}
\end{algorithm}

\section{Simulation Study}
\label{sec:simulation}

We evaluate the proposed framework using sparse SIR, SAVE, and
both $y$-based and $r$-based PHD, focusing on recovery of the true
dimension reduction subspace under block missingness.
Results for three dimension reduction methods and their
corresponding response models are presented below, with the
remaining results reported in
Appendix~\ref{app:additional-simulations}.

\subsection{Simulation Design}
\label{sec:simulation-design}

\paragraph{Data generation and missingness.}
We generate $\mathbf{X}\sim N_{30}(\mathbf{0},\boldsymbol{\Sigma})$
with three modalities labeled A, B, and C, each containing
ten predictors. The target
dimension is $d=2$, and the true dimension reduction subspace
is spanned by the columns of
$\mathbf{B}_0=(\boldsymbol{\beta}_1,\boldsymbol{\beta}_2)$.
The nonzero entries of $\boldsymbol{\beta}_1$ are
$1/\sqrt{6}$ on coordinates $1$--$6$ in modality A,
and those of $\boldsymbol{\beta}_2$ are $1/\sqrt{6}$
on coordinates $11$--$13$ in modality B and $21$--$23$
in modality C. Write
$\boldsymbol{\eta}=(\eta_1,\eta_2)^\top
=\mathbf{B}_0^\top\mathbf{X}$ and let
$\epsilon\sim N(0,1)$ be independent of $\mathbf{X}$.

Each replication contains 400 training, 200 validation, and
200 test observations. Validation and test observations are
complete, while the training data exhibit block missingness.
All patterns retain the same 100 complete training observations,
while P2 and P3 successively reveal additional observations
of modality A.

\begin{figure}[!htb]
\centering
\setstretch{1}
\includegraphics[width=\textwidth]{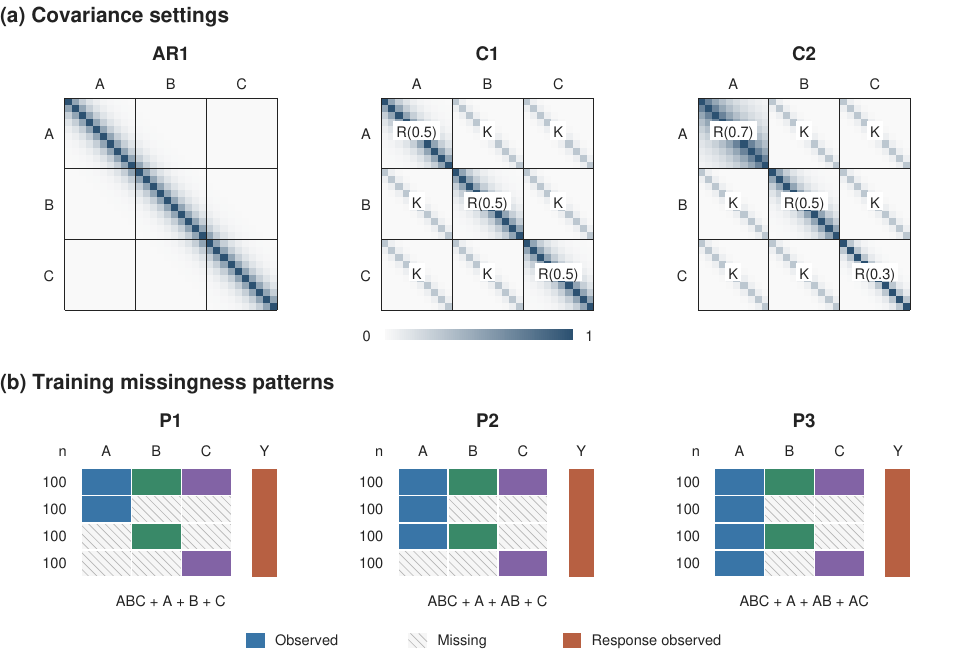}
\captionsetup{font=small}
\caption{Simulation settings.}
\label{fig:simulation-design}
\end{figure}

Figure~\ref{fig:simulation-design} shows the three covariance
settings and three training missingness patterns.
In panel (a), let
$\mathbf{R}_m(\rho)=(\rho^{|j-t|})_{j,t=1}^m$.
AR1 uses $\mathbf{R}_{30}(0.5)$, while C1 and C2 have the
displayed block structures, where
$\mathbf{R}(\rho)=\mathbf{R}_{10}(\rho)$ and
$\mathbf{K}=0.3\mathbf{R}_{10}(0.2)$.
Panel (b) shows four groups of 100 subjects aligned across
P1--P3. Colored blocks are observed and hatched blocks
are missing.

\paragraph{Competing estimators and evaluation.}
We compare complete-case, available-case, and SoftImpute-based
\citep{mazumder2010spectral} estimators with the proposed
shrinkage estimator using a common fitting and tuning procedure
based on Algorithm~\ref{alg:block-missing-dr}.
To construct the method-specific matrices, complete-case
estimation uses fully observed subjects, available-case
estimation uses all observations available for each required
moment, and SoftImpute completes the predictor matrix using
a low-rank approximation before moment estimation.
All estimators require tuning the solver penalty parameters.
For SoftImpute, the additional regularization parameter is
tuned jointly with the solver penalties within the algorithm.
We fix the number of slices at $H=8$ for SIR and SAVE.

Each setting uses 100 replications, with shared data and
missingness patterns across estimators. We measure how well
each estimator recovers the true sufficient subspace using
the projection distance
$\Delta(\widehat{\mathbf{B}},\mathbf{B}_0)
=\|\widehat{\mathbf{Q}}\widehat{\mathbf{Q}}^\top
-\mathbf{B}_0\mathbf{B}_0^\top\|_F$,
where $\widehat{\mathbf{Q}}$ is an orthonormal basis of
$\operatorname{col}(\widehat{\mathbf{B}})$, and
$\mathbf{B}_0$ has orthonormal columns by construction.

\subsection{Subspace Recovery Results}
\label{sec:simulation-results}

\begin{figure}[!htbp]
\centering
\setstretch{1}
\includegraphics[width=\textwidth]{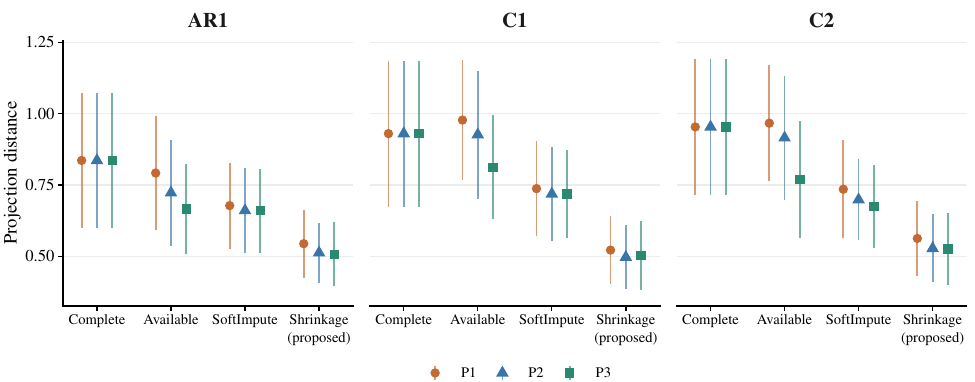}
\captionsetup{font=small}
\caption{SIR with the rational response.}
\label{fig:simulation-sir}
\end{figure}

\begin{figure}[!htbp]
\centering
\setstretch{1}
\includegraphics[width=\textwidth]{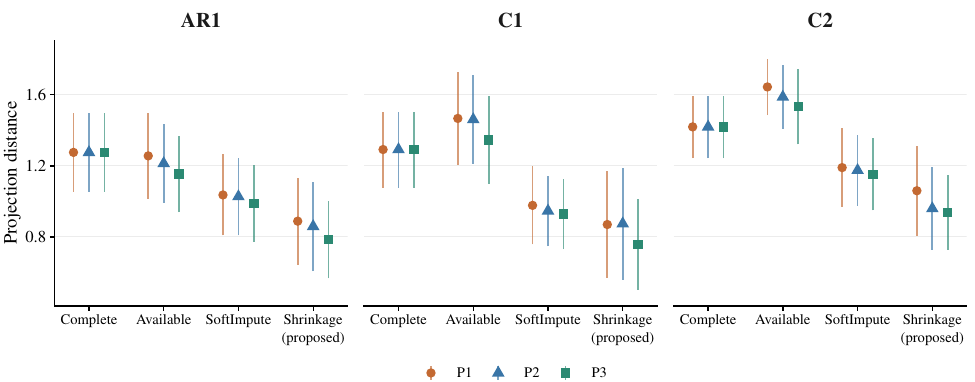}
\captionsetup{font=small}
\caption{SAVE with the quadratic--absolute response.}
\label{fig:simulation-save}
\end{figure}

\begin{figure}[!htbp]
\centering
\setstretch{1}
\includegraphics[width=\textwidth]{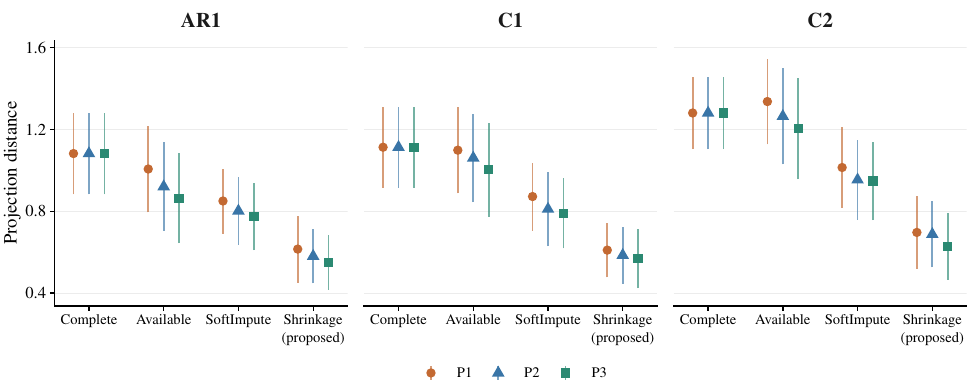}
\captionsetup{font=small}
\caption{$y$-based PHD with the radial logarithmic response.}
\label{fig:simulation-phd}
\end{figure}

Figures~\ref{fig:simulation-sir}--\ref{fig:simulation-phd}
show mean projection distance $\pm$ SD over successful fits
for the methods and response models specified below.
The three panels correspond to covariance settings AR1,
C1, and C2, while colors and symbols distinguish P1--P3.
Shrinkage achieves the lowest mean projection distance
in all 27 settings.

Mean recovery error typically
rises as missingness increases from P3 to P1 and the covariance
setting changes from AR1 to C2. Across these changes,
shrinkage generally shows a smaller increase in recovery
error than available-case estimation. These results demonstrate
improved subspace recovery and support the robustness of
the proposed method across the examined covariance settings
and block-missingness patterns.

\paragraph{SIR.}
We use the rational response
$Y=\eta_1/\{0.5+(\eta_2+1.5)^2\}+0.2\epsilon$.

\newpage

\paragraph{SAVE.}
We use the quadratic--absolute response
$Y=\eta_1^2+2|\eta_2|+0.1|\eta_2|\epsilon$.

\paragraph{$\boldsymbol{y}$-based PHD.}
We use the radial logarithmic response
$Y=r\log r+0.2\epsilon$, where
$r=(\eta_1^2+\eta_2^2)^{1/2}$ and $r\log r=0$ at $r=0$.

The additional simulation results for SIR, SAVE, and
$r$-based PHD in Appendix~\ref{app:additional-simulations}
also show that the proposed estimator achieves the lowest
mean projection distance in all 27 settings.

\newpage

\subsection{Secondary Results}
\label{sec:simulation-secondary}

Tables~\ref{tab:simulation-secondary-ar1p1}--\ref{tab:simulation-secondary-c2p1}
summarize secondary results under missingness pattern P1
across the AR1, C1, and C2 covariance settings.
The success rate is the percentage of replications yielding
a valid fit.
LOO-KNN MSE measures prediction error in the estimated
subspace on a separate test set, using the same LOO-KNN
procedure as in tuning.
After matching estimated loadings to the true loadings using
absolute cosine similarity, we denote the numbers of
selected relevant variables by $\mathrm{TP}_1/\mathrm{TP}_2$,
out of six for each loading.
$\mathrm{FP}_1/\mathrm{FP}_2$ denote the corresponding
numbers of irrelevant variables selected.

Available-case failures arise from loss of positive
definiteness and are especially frequent under substantial
missingness with poorly conditioned covariance.
The proposed estimator generally recovers more relevant
variables, although its loadings are not typically sparser
than those of other methods.
It also gives the lowest LOO-KNN MSE for SIR and $y$-based PHD.
For SAVE, SoftImpute gives lower MSE despite poorer
subspace recovery.

\input{tb_simulation_secondary}

\newpage

\section{Real data analysis}
\label{sec:real-data}

\subsection{Data and Preprocessing}
\label{sec:adni-data}

The Alzheimer's Disease Neuroimaging Initiative (ADNI) is a
longitudinal study collecting imaging and genetic data,
biofluid samples, and clinical assessments to investigate
Alzheimer's disease (AD) progression and thereby support early detection
and the treatment development for AD \citep{weiner2017adni3}.
In this study, we focus on the multimodal nature of the data,
using predictors from three modalities: magnetic resonance
imaging (MRI), positron emission tomography (PET), and
cerebrospinal fluid (CSF) biomarkers.
The response is the 13-item Alzheimer's Disease Assessment
Scale--Cognitive Subscale (ADAS-Cog13), which assesses
impairment in cognitive domains affected by AD, with higher
scores indicating poorer cognitive performance
\citep{mohs1997development}.

We include 74 predictors: 34 each from MRI and tau PET,
and six CSF biomarkers.
The MRI and tau PET images are processed to obtain regional
cortical thickness and tracer uptake measures, respectively.
For both imaging modalities, we use the Desikan--Killiany atlas,
which divides each cerebral hemisphere into 34 anatomically
defined cortical regions \citep{desikan2006automated}.
We average measurements from corresponding left and right
hemisphere regions to obtain one predictor per region for
each modality.
The CSF predictors are measurements of $A\beta40$, $A\beta42$,
total tau, and phosphorylated tau at threonine 181.
We also include total tau/$A\beta42$ and phosphorylated
tau/$A\beta42$ to represent relative levels of tau and amyloid
biomarkers, as in \citet{thung2014neurodegenerative}.

\begin{figure}[htbp]
\centering
\includegraphics[
  width=0.8\textwidth,
  height=0.4\textheight,
  keepaspectratio
]{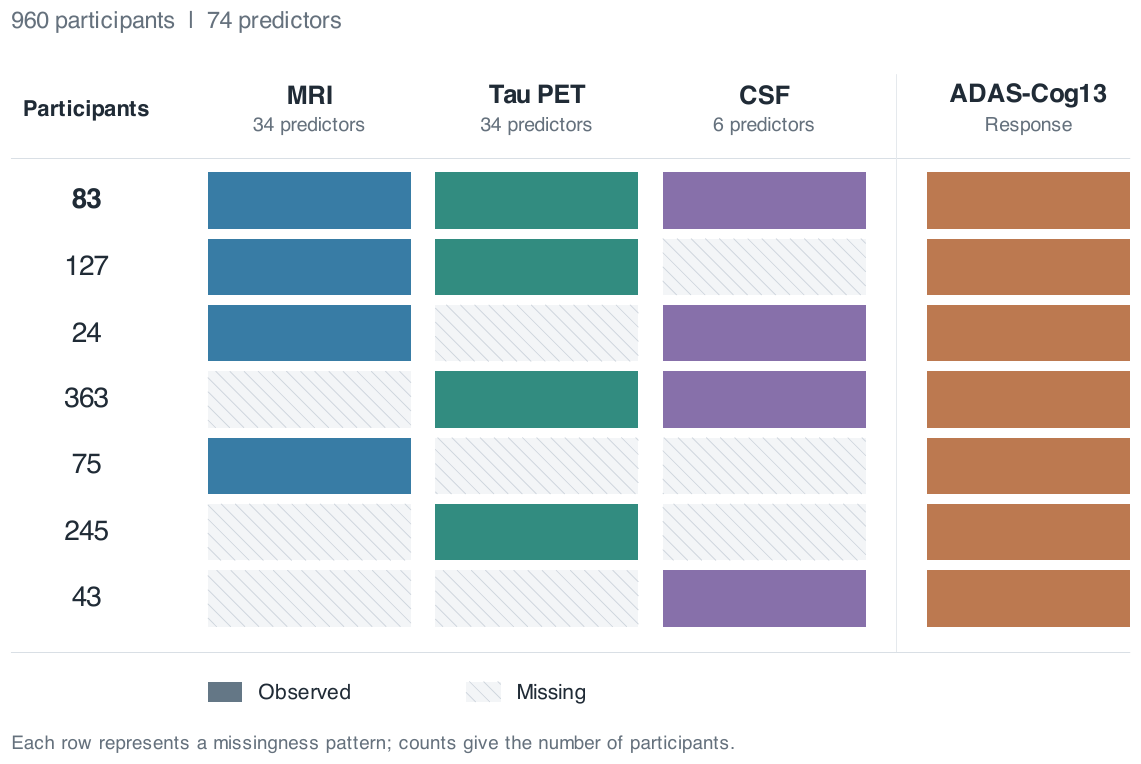}
\caption{Block-missingness patterns in the ADNI3 data.}
\label{fig:adni-missingness}
\end{figure}

We combine the three modalities and the response using participant
IDs and visit information. For consistency, we retain observations
with an observed response from the ADNI3 phase and select one
assessment per participant at their first ADNI3 visit.
Participants may lack measurements for one or more modalities,
resulting in the block-missingness patterns shown in
Figure~\ref{fig:adni-missingness}.

We apply Algorithm~\ref{alg:block-missing-dr} to these
multimodal block-missing data. We randomly allocate
50 complete participants to the validation set for tuning.
The remaining 33 complete participants and all
877 incomplete participants form the training set.
For each split, we standardize each predictor using its
observed training values and fit sparse SIR within
the proposed framework with $d=2$.

We compare shrinkage with complete-case, available-case,
and SoftImpute-based estimators using the same fitting
and tuning procedure as in the simulation studies.
All estimators use the same training/validation splits.
Under this allocation, complete-case estimation reduces
the training sample to 33 participants, so its covariance
estimate for 74 predictors is necessarily rank-deficient.

\subsection{Data Analysis Results}
\label{sec:adni-results}

\begin{figure}[tbp]
\centering
\begin{minipage}[c]{0.60\linewidth}
\includegraphics[width=\linewidth]{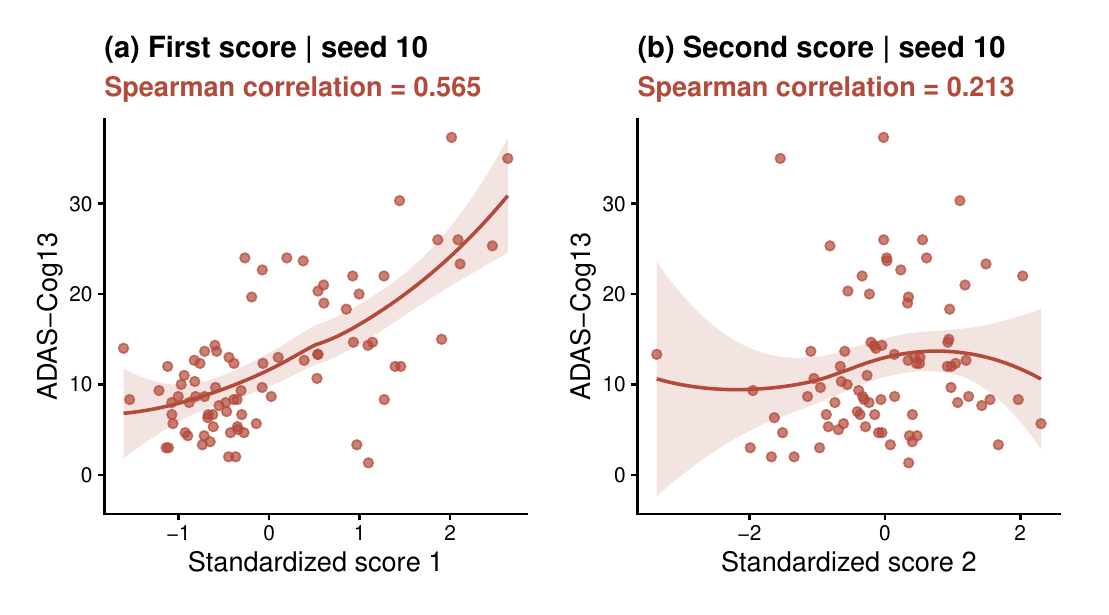}
\end{minipage}\hfill
\begin{minipage}[c]{0.39\linewidth}
\includegraphics[width=\linewidth]{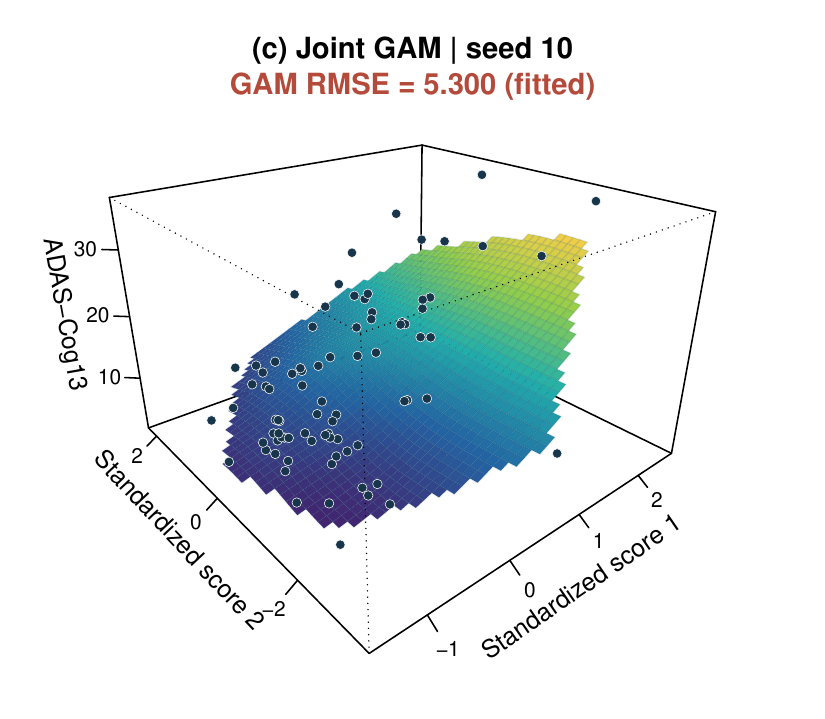}
\end{minipage}
\caption{Illustrative scatterplots showing all 83 complete participants.}
\label{fig:adni-scores}
\end{figure}

To examine the associations retained by the estimated dimension
reduction subspace, we plot ADAS-Cog13 against the projected
scores separately and jointly.
We overlay fitted LOESS curves for the individual scores
and a fitted generalized additive model (GAM) surface for both
scores to visualize these relationships. Figure~\ref{fig:adni-scores} presents an illustrative fit
of sparse SIR with the shrinkage estimator, showing all 83 complete participants.

Table~\ref{tab:adni-results} compares the four estimators
across 100 random training/validation splits.
Success reports the number of successful fits.
The absolute Spearman correlations summarize monotonic
associations between the response and each projected score,
ordered from larger to smaller.
GAM RMSE measures how closely the joint GAM fitted to the
two scores describes the responses among the 83 complete
participants.
Selected denotes the number of predictors selected in
at least one of the two loadings, out of the 74 predictors.

\begin{table}[htbp]
\centering
{\input{table_results.tex}}
\caption{Sparse SIR results over 100 random allocations:
mean (SD), except for Success.}
\label{tab:adni-results}
\end{table}

Complete-case and available-case estimation fail in all
100 replications because their covariance estimates
are not positive definite.
Compared with SoftImpute, the proposed estimator yields
higher mean absolute Spearman correlations, particularly
for the second direction, and a lower mean fitted GAM RMSE.
Shrinkage selects more predictors on average, yielding
a less sparse representation. Note that the correlations and RMSE are descriptive summaries
of the associations retained by the estimated subspace,
rather than independent test-set performance measures.

To further understand the results, we examine five predictors
selected in at least one loading in all 100 random allocations:
entorhinal and transverse temporal cortical thickness,
entorhinal and fusiform tau PET uptake, and the CSF
total tau/$A\beta42$ ratio.
If we orient each loading vector so that its corresponding
score has a positive Spearman correlation with ADAS-Cog13,
the nonzero coefficients are consistently negative for
the MRI predictors and positive for the PET predictors
and CSF ratio.
These findings are broadly consistent with reports of
cortical thinning in AD and associations of elevated
tau-related biomarkers with poorer cognition or
subsequent clinical decline
\citep{li2025voxel,ossenkoppele2021tau,hansson2018csf}. 

Overall, the ADNI analysis illustrates the applicability
of the proposed framework in three respects.
First, it incorporates all participants without imputation
and ensures positive-definite covariance estimates.
Second, it produces low-dimensional scores associated
with ADAS-Cog13, achieving higher mean absolute Spearman
correlations and lower mean fitted GAM RMSE than SoftImpute.
The resulting representations are less sparse, and
a sparser alternative can be explored by increasing
$\lambda_1$ after hyperparameter selection.
Third, the consistently selected predictors and their
loading signs provide a biologically plausible
interpretation of the estimated representation.

\section{Conclusion}
\label{sec:conclusion}

We proposed a sparse sufficient dimension reduction
framework for block-missing multimodal data. The method
applies separate shrinkage to within- and cross-modality
covariance components to construct method-specific matrices,
allowing SIR, SAVE, and PHD to be fitted through a common
sparse solver. We also proposed a prediction-based criterion
for jointly selecting the shrinkage and solver penalty
parameters. Simulation studies demonstrated consistent
improvements in subspace recovery across the examined
response models, covariance structures, and missingness
patterns.

The ADNI analysis illustrated the method's application to
multimodal clinical data and raised further questions.
Future directions include adapting shrinkage to differences
in modality availability and dimensionality, exploiting
relationships among tuning parameters to improve computational
efficiency in higher-dimensional settings, and refining
method-specific kernel estimation. These extensions can
build on the proposed framework while retaining its common
fitting and tuning structure.

\clearpage
\appendix
\input{appendix}

\clearpage
\bibliographystyle{plainnat}
\bibliography{references}

\end{document}

%% file: head.tex
\usepackage{graphicx}
\usepackage{mathrsfs}
\usepackage{amsthm,amssymb,amsmath,dsfont, bbm, mathtools, amsfonts, amsthm}
\usepackage{natbib} 
\usepackage{enumerate}
\usepackage{array}
\usepackage{booktabs}
\usepackage{adjustbox}
\usepackage[normalem]{ulem} 
\usepackage{xcolor}
\usepackage[margin = 1.0in]{geometry}
\usepackage{setspace}
\usepackage{mathabx}
\usepackage{multirow,verbatim}
\usepackage{indentfirst}
\usepackage[english]{babel}
\usepackage{algorithm}
\usepackage{algpseudocode}
\usepackage{enumitem}
\usepackage{hyperref}
\usepackage{caption}

\hypersetup{
colorlinks=true,
citecolor=blue,
linkcolor=blue,
urlcolor=blue
}

\usepackage[pagewise, mathlines]{lineno}

\newfont{\rsfsten}{rsfs10 scaled 1100}
\newfont{\rsfstena}{rsfs10 scaled 800}
\newfont{\rsfstenb}{rsfs10 scaled 500}

\theoremstyle{plain}
\newtheorem{theorem}{Theorem}

\newtheorem{lemma}[theorem]{Lemma}

\theoremstyle{definition}

\theoremstyle{remark}

\newcommand{\bB}{\mathbf{B}}

\newcommand{\bbeta}{\boldsymbol{\beta}}

\newcommand{\bphi}{\boldsymbol{\phi}}

\newcommand{\benu}{\begin{enumerate}}
\newcommand{\eenu}{\end{enumerate}}
\newcommand{\bi}{\begin{itemize}}
\newcommand{\ei}{\end{itemize}}
\newcommand{\beq}{\begin{eqnarray*}}
\newcommand{\eeq}{\end{eqnarray*}}
\newcommand{\bbox}{\begin{tcolorbox}}
\newcommand{\ebox}{\end{tcolorbox}}

%% file: tb_dr_list.tex
\begin{table}[!htbp]
\centering
\setstretch{1}
\small
\renewcommand{\arraystretch}{1.2}
\begin{tabular*}{\textwidth}{@{\extracolsep{\fill}}lll@{}}
\toprule
Method & $\mathbf{M}$ & $\mathbf{G}$ \\
\midrule
PCA & $\boldsymbol{\Sigma}$ & $\mathbf{I}_p$ \\
SIR
& $\boldsymbol{\Sigma}^{1/2}
   \operatorname{Cov}\{E(\mathbf{Z}\mid Y)\}
   \boldsymbol{\Sigma}^{1/2}$
& $\boldsymbol{\Sigma}$ \\
SAVE & $\boldsymbol{\Sigma}^{1/2}E[\{\mathbf{I}_p-\operatorname{Cov}(\mathbf{Z}\mid Y)\}^{2}]\boldsymbol{\Sigma}^{1/2}$ & $\boldsymbol{\Sigma}$ \\
PHD ($y$-based) & $\boldsymbol{\Sigma}^{1/2}\boldsymbol{\Sigma}_{YZZ}^{2}\boldsymbol{\Sigma}^{1/2}$ & $\boldsymbol{\Sigma}$ \\
PHD ($r$-based) & $\boldsymbol{\Sigma}^{1/2}\boldsymbol{\Sigma}_{rZZ}^{2}\boldsymbol{\Sigma}^{1/2}$ & $\boldsymbol{\Sigma}$ \\
\bottomrule
\end{tabular*}
\caption{Population matrix pairs for dimension reduction methods.}
\label{tab:generalized-eigen}
\end{table}

%% file: tb_estimated_pairs.tex
\begin{table}[!htbp]
\centering
\setstretch{1}
\small
\renewcommand{\arraystretch}{1.2}
\begin{tabular*}{\textwidth}{@{\extracolsep{\fill}}lll@{}}
\toprule
Method & $\widehat{\mathbf{M}}$ & $\widehat{\mathbf{G}}$ \\
\midrule
SIR
& $\displaystyle\sum_{h=1}^{H}\widehat p_h
(\widetilde{\boldsymbol{\mu}}_{(h)}-\widetilde{\boldsymbol{\mu}})
(\widetilde{\boldsymbol{\mu}}_{(h)}-\widetilde{\boldsymbol{\mu}})^\top$
& $\widehat{\boldsymbol{\Sigma}}$ \\
\addlinespace
SAVE
& $\displaystyle\sum_{h=1}^{H}\widehat p_h
(\widehat{\boldsymbol{\Sigma}}-\widetilde{\boldsymbol{\Gamma}}_{(h)})
\widehat{\boldsymbol{\Sigma}}^{-1}
(\widehat{\boldsymbol{\Sigma}}-\widetilde{\boldsymbol{\Gamma}}_{(h)})$
& $\widehat{\boldsymbol{\Sigma}}$ \\
\addlinespace
PHD ($y$-based)
& $\widetilde{\mathbf{T}}_Y
\widehat{\boldsymbol{\Sigma}}^{-1}\widetilde{\mathbf{T}}_Y$
& $\widehat{\boldsymbol{\Sigma}}$ \\
\addlinespace
PHD ($r$-based)
& $\widetilde{\mathbf{T}}_r
\widehat{\boldsymbol{\Sigma}}^{-1}\widetilde{\mathbf{T}}_r$
& $\widehat{\boldsymbol{\Sigma}}$ \\
\addlinespace
PCA
& $\widehat{\boldsymbol{\Sigma}}$
& $\mathbf{I}_p$ \\
\bottomrule
\end{tabular*}
\caption{Estimated matrix pairs under block missingness.}
\label{tab:estimated-pairs}
\end{table}

%% file: tb_simulation_secondary.tex

\begin{table}[!htbp]
\centering
\setstretch{1}
\small
\renewcommand{\arraystretch}{1.1}
\begin{tabular*}{\textwidth}{@{\extracolsep{\fill}}lrrrr@{}}
\toprule
Estimator & Success rate (\%) & LOO-KNN MSE & $\mathrm{TP}_1/\mathrm{TP}_2$ & $\mathrm{FP}_1/\mathrm{FP}_2$ \\
\midrule
\multicolumn{5}{@{}l}{\textit{SIR: rational}} \\
Complete & \textbf{100} & 0.508 (0.153) & 5.95/5.48 & 1.08/1.99 \\
Available & \textbf{100} & 0.485 (0.142) & 5.98/5.65 & \textbf{0.47}/\textbf{0.93} \\
SoftImpute & \textbf{100} & 0.449 (0.133) & \textbf{6.00}/5.95 & 1.51/3.13 \\
Shrinkage (proposed) & \textbf{100} & \textbf{0.434} (0.139) & 5.99/\textbf{6.00} & 1.90/3.64 \\
\midrule
\multicolumn{5}{@{}l}{\textit{SAVE: quadratic--absolute}} \\
Complete & \textbf{100} & 6.693 (1.949) & 5.47/3.73 & \textbf{2.11}/\textbf{2.46} \\
Available & \textbf{100} & 6.656 (1.972) & 5.60/4.16 & 4.76/4.86 \\
SoftImpute & \textbf{100} & \textbf{5.782} (1.780) & 5.93/4.87 & 4.24/4.64 \\
Shrinkage (proposed) & \textbf{100} & 5.784 (1.819) & \textbf{5.99}/\textbf{5.73} & 3.08/6.38 \\
\midrule
\multicolumn{5}{@{}l}{\textit{$y$-based PHD: radial logarithmic}} \\
Complete & \textbf{100} & 1.233 (0.283) & 5.46/5.31 & 2.82/2.82 \\
Available & \textbf{100} & 1.180 (0.277) & 5.52/5.51 & \textbf{2.13}/\textbf{2.55} \\
SoftImpute & \textbf{100} & 1.060 (0.212) & 5.92/5.85 & 4.21/4.80 \\
Shrinkage (proposed) & \textbf{100} & \textbf{0.959} (0.195) & \textbf{5.99}/\textbf{5.96} & 2.71/4.34 \\
\bottomrule
\end{tabular*}
\caption{Secondary results under AR1--P1.
MSE is reported as mean (SD).}
\label{tab:simulation-secondary-ar1p1}
\end{table}

\begin{table}[!htbp]
\centering
\setstretch{1}
\small
\renewcommand{\arraystretch}{1.1}
\begin{tabular*}{\textwidth}{@{\extracolsep{\fill}}lrrrr@{}}
\toprule
Estimator & Success rate (\%) & LOO-KNN MSE & $\mathrm{TP}_1/\mathrm{TP}_2$ & $\mathrm{FP}_1/\mathrm{FP}_2$ \\
\midrule
\multicolumn{5}{@{}l}{\textit{SIR: rational}} \\
Complete & \textbf{100} & 0.542 (0.165) & 5.91/5.11 & 0.57/1.31 \\
Available & 76 & 0.551 (0.143) & 5.96/4.87 & \textbf{0.32}/\textbf{0.51} \\
SoftImpute & \textbf{100} & 0.469 (0.142) & 5.99/5.85 & 1.80/2.86 \\
Shrinkage (proposed) & \textbf{100} & \textbf{0.437} (0.127) & \textbf{6.00}/\textbf{6.00} & 2.80/3.37 \\
\midrule
\multicolumn{5}{@{}l}{\textit{SAVE: quadratic--absolute}} \\
Complete & \textbf{100} & 7.163 (2.243) & 5.45/3.75 & \textbf{1.96}/\textbf{1.71} \\
Available & 80 & 7.708 (2.498) & 5.47/4.05 & 8.26/8.84 \\
SoftImpute & \textbf{100} & \textbf{5.923} (1.861) & 5.92/5.11 & 3.35/3.20 \\
Shrinkage (proposed) & \textbf{100} & 6.178 (2.219) & \textbf{5.93}/\textbf{5.57} & 5.95/6.99 \\
\midrule
\multicolumn{5}{@{}l}{\textit{$y$-based PHD: radial logarithmic}} \\
Complete & \textbf{100} & 1.409 (0.283) & 5.44/5.37 & \textbf{3.16}/\textbf{3.38} \\
Available & 78 & 1.391 (0.321) & 5.51/5.47 & 4.97/5.28 \\
SoftImpute & \textbf{100} & 1.201 (0.268) & 5.93/5.87 & 4.70/5.07 \\
Shrinkage (proposed) & \textbf{100} & \textbf{1.090} (0.231) & \textbf{5.98}/\textbf{6.00} & 3.95/5.07 \\
\bottomrule
\end{tabular*}
\caption{Secondary results under C1--P1.}
\end{table}

\begin{table}[!htbp]
\centering
\setstretch{1}
\small
\renewcommand{\arraystretch}{1.1}
\begin{tabular*}{\textwidth}{@{\extracolsep{\fill}}lrrrr@{}}
\toprule
Estimator & Success rate (\%) & LOO-KNN MSE & $\mathrm{TP}_1/\mathrm{TP}_2$ & $\mathrm{FP}_1/\mathrm{FP}_2$ \\
\midrule
\multicolumn{5}{@{}l}{\textit{SIR: rational}} \\
Complete & \textbf{100} & 0.757 (0.241) & 5.77/5.23 & 0.73/1.25 \\
Available & 7 & 0.910 (0.149) & 5.86/5.57 & \textbf{0.57}/\textbf{0.43} \\
SoftImpute & \textbf{100} & 0.645 (0.195) & 5.95/5.93 & 1.72/2.65 \\
Shrinkage (proposed) & \textbf{100} & \textbf{0.633} (0.208) & \textbf{5.99}/\textbf{5.96} & 3.22/2.65 \\
\midrule
\multicolumn{5}{@{}l}{\textit{SAVE: quadratic--absolute}} \\
Complete & \textbf{100} & 10.644 (3.538) & 5.15/2.75 & \textbf{1.35}/\textbf{1.83} \\
Available & 10 & 12.678 (6.758) & 5.00/3.00 & 6.30/8.80 \\
SoftImpute & \textbf{100} & \textbf{9.554} (3.193) & 5.78/4.40 & 3.13/3.71 \\
Shrinkage (proposed) & \textbf{100} & 9.820 (3.463) & \textbf{5.97}/\textbf{4.74} & 4.14/5.60 \\
\midrule
\multicolumn{5}{@{}l}{\textit{$y$-based PHD: radial logarithmic}} \\
Complete & \textbf{100} & 1.843 (0.393) & 4.92/5.09 & 2.65/\textbf{3.91} \\
Available & 13 & 1.748 (0.378) & 5.00/4.31 & \textbf{2.62}/5.38 \\
SoftImpute & \textbf{100} & 1.513 (0.320) & 5.69/5.73 & 4.33/5.14 \\
Shrinkage (proposed) & \textbf{100} & \textbf{1.362} (0.292) & \textbf{5.96}/\textbf{5.96} & 4.04/4.47 \\
\bottomrule
\end{tabular*}
\caption{Secondary results under C2--P1.}
\label{tab:simulation-secondary-c2p1}
\label{tab:simulation-secondary}
\end{table}

%% file: table_results.tex
\small
\setlength{\tabcolsep}{3pt}
\renewcommand{\arraystretch}{1.15}
\begin{tabular*}{\linewidth}{@{\extracolsep{\fill}}lrrrrr@{}}
\toprule
Method & Success & Spearman $|\rho_1|$ & Spearman $|\rho_2|$ & GAM RMSE & Selected \\
\midrule
Complete-case & 0/100 & --- & --- & --- & --- \\
Available-case & 0/100 & --- & --- & --- & --- \\
SoftImpute & \textbf{100/100} & 0.507 (0.017) & 0.093 (0.040) & 5.691 (0.288) & \textbf{18.0 (4.4)} \\
Shrinkage (proposed) & \textbf{100/100} & \textbf{0.528 (0.035)} & \textbf{0.271 (0.066)} & \textbf{5.500 (0.335)} & 38.3 (12.4) \\
\bottomrule
\end{tabular*}

%% file: appendix.tex
\section{Method-Specific Kernel Construction}
\label{app:matrix-derivations}

We construct method-specific kernel matrices using the shrinkage
covariance estimator and available-case moments.
In Section~\ref{app:population-kernels}, we express the population
kernels in Table~\ref{tab:generalized-eigen} in the original
coordinates using
\begin{equation}
\label{eq:standardized-predictor}
\mathbf{Z}
=\boldsymbol{\Sigma}^{-1/2}(\mathbf{X}-\boldsymbol{\mu}),
\end{equation}
to avoid directly whitening incomplete observations.
In Section~\ref{app:available-case-moments}, we replace
$\boldsymbol{\Sigma}$ with $\widehat{\boldsymbol{\Sigma}}$
and estimate the remaining moments using all observations
available for each moment.

\subsection{Population kernel expressions}
\label{app:population-kernels}

\paragraph{SIR.}
Denote the SIR kernel by
$\mathbf{M}_{\mathrm{SIR}}
=\boldsymbol{\Sigma}^{1/2}
\operatorname{Cov}\{E(\mathbf{Z}\mid Y)\}
\boldsymbol{\Sigma}^{1/2}$.
Using \eqref{eq:standardized-predictor}, we obtain
\[
\begin{aligned}
E(\mathbf{Z}\mid Y)
&=E\{\boldsymbol{\Sigma}^{-1/2}
(\mathbf{X}-\boldsymbol{\mu})\mid Y\}\\
&=\boldsymbol{\Sigma}^{-1/2}
\{E(\mathbf{X}\mid Y)-\boldsymbol{\mu}\}.
\end{aligned}
\]
Since $\boldsymbol{\mu}$ and $\boldsymbol{\Sigma}$ are constant, $\operatorname{Cov}\{E(\mathbf{Z}\mid Y)\}=\boldsymbol{\Sigma}^{-1/2}\operatorname{Cov}\{E(\mathbf{X}\mid Y)\}\boldsymbol{\Sigma}^{-1/2}$. 
Thus,
\[
\begin{aligned}
\mathbf{M}_{\mathrm{SIR}}
&=\boldsymbol{\Sigma}^{1/2}
\operatorname{Cov}\{E(\mathbf{Z}\mid Y)\}
\boldsymbol{\Sigma}^{1/2}\\
&=\operatorname{Cov}\{E(\mathbf{X}\mid Y)\}.
\end{aligned}
\]

\paragraph{SAVE.}
Denote the SAVE kernel by
$\mathbf{M}_{\mathrm{SAVE}}
=\boldsymbol{\Sigma}^{1/2}
E[\{\mathbf{I}_p-\operatorname{Cov}(\mathbf{Z}\mid Y)\}^{2}]
\boldsymbol{\Sigma}^{1/2}$.
Let $\boldsymbol{\Gamma}(Y)=\operatorname{Cov}(\mathbf{X}\mid Y)$.
Using \eqref{eq:standardized-predictor}, we obtain
\[
\begin{aligned}
\operatorname{Cov}(\mathbf{Z}\mid Y)
&=\operatorname{Cov}\{
\boldsymbol{\Sigma}^{-1/2}(\mathbf{X}-\boldsymbol{\mu})\mid Y\}\\
&=\boldsymbol{\Sigma}^{-1/2}
\boldsymbol{\Gamma}(Y)\boldsymbol{\Sigma}^{-1/2}.
\end{aligned}
\]
Since $\mathbf{I}_p=\boldsymbol{\Sigma}^{-1/2}\boldsymbol{\Sigma}\boldsymbol{\Sigma}^{-1/2}$, $\mathbf{I}_p-\operatorname{Cov}(\mathbf{Z}\mid Y)=\boldsymbol{\Sigma}^{-1/2}\{\boldsymbol{\Sigma}-\boldsymbol{\Gamma}(Y)\}\boldsymbol{\Sigma}^{-1/2}$.
Thus,
\[
\begin{aligned}
\mathbf{M}_{\mathrm{SAVE}}
&=\boldsymbol{\Sigma}^{1/2}
E[\{\mathbf{I}_p-\operatorname{Cov}(\mathbf{Z}\mid Y)\}^{2}]
\boldsymbol{\Sigma}^{1/2}\\
&=\boldsymbol{\Sigma}^{1/2}
E\!\left[
\boldsymbol{\Sigma}^{-1/2}
\{\boldsymbol{\Sigma}-\boldsymbol{\Gamma}(Y)\}
\boldsymbol{\Sigma}^{-1}
\{\boldsymbol{\Sigma}-\boldsymbol{\Gamma}(Y)\}
\boldsymbol{\Sigma}^{-1/2}
\right]
\boldsymbol{\Sigma}^{1/2}\\
&=E\!\left[
\{\boldsymbol{\Sigma}-\boldsymbol{\Gamma}(Y)\}
\boldsymbol{\Sigma}^{-1}
\{\boldsymbol{\Sigma}-\boldsymbol{\Gamma}(Y)\}
\right].
\end{aligned}
\]

\paragraph{$\boldsymbol{y}$-based PHD.}
Denote the $y$-based PHD kernel by
$\mathbf{M}_{\mathrm{PHD},y}
=\boldsymbol{\Sigma}^{1/2}
\boldsymbol{\Sigma}_{YZZ}^{2}\boldsymbol{\Sigma}^{1/2}$,
where
$\boldsymbol{\Sigma}_{YZZ}
=E[\{Y-E(Y)\}\mathbf{Z}\mathbf{Z}^\top]$.
Let $\mathbf{T}_Y
=E[\{Y-E(Y)\}(\mathbf{X}-\boldsymbol{\mu})
(\mathbf{X}-\boldsymbol{\mu})^\top]$.
Using \eqref{eq:standardized-predictor}, we obtain
\[
\begin{aligned}
\boldsymbol{\Sigma}_{YZZ}
&=E[\{Y-E(Y)\}\mathbf{Z}\mathbf{Z}^\top]\\
&=E\!\left[
\{Y-E(Y)\}\boldsymbol{\Sigma}^{-1/2}
(\mathbf{X}-\boldsymbol{\mu})
(\mathbf{X}-\boldsymbol{\mu})^\top
\boldsymbol{\Sigma}^{-1/2}
\right]\\
&=E\!\left[
\boldsymbol{\Sigma}^{-1/2}\{Y-E(Y)\}
(\mathbf{X}-\boldsymbol{\mu})
(\mathbf{X}-\boldsymbol{\mu})^\top
\boldsymbol{\Sigma}^{-1/2}
\right]
\quad\text{(since $Y-E(Y)$ is scalar)}\\
&=\boldsymbol{\Sigma}^{-1/2}
E[\{Y-E(Y)\}(\mathbf{X}-\boldsymbol{\mu})
(\mathbf{X}-\boldsymbol{\mu})^\top]
\boldsymbol{\Sigma}^{-1/2}\\
&=\boldsymbol{\Sigma}^{-1/2}\mathbf{T}_Y
\boldsymbol{\Sigma}^{-1/2}.
\end{aligned}
\]
Thus,
\[
\begin{aligned}
\mathbf{M}_{\mathrm{PHD},y}
&=\boldsymbol{\Sigma}^{1/2}
\boldsymbol{\Sigma}_{YZZ}^{2}\boldsymbol{\Sigma}^{1/2}\\
&=\boldsymbol{\Sigma}^{1/2}
\bigl\{\boldsymbol{\Sigma}^{-1/2}\mathbf{T}_Y
\boldsymbol{\Sigma}^{-1}\mathbf{T}_Y
\boldsymbol{\Sigma}^{-1/2}\bigr\}
\boldsymbol{\Sigma}^{1/2}\\
&=\mathbf{T}_Y\boldsymbol{\Sigma}^{-1}\mathbf{T}_Y.
\end{aligned}
\]

\paragraph{$\boldsymbol{r}$-based PHD.}
Denote the $r$-based PHD kernel by
$\mathbf{M}_{\mathrm{PHD},r}
=\boldsymbol{\Sigma}^{1/2}
\boldsymbol{\Sigma}_{rZZ}^{2}\boldsymbol{\Sigma}^{1/2}$,
where $\boldsymbol{\Sigma}_{rZZ}=E(r\mathbf{Z}\mathbf{Z}^\top)$.
Let $\mathbf{T}_r
=E[r(\mathbf{X}-\boldsymbol{\mu})
(\mathbf{X}-\boldsymbol{\mu})^\top]$.
Using \eqref{eq:standardized-predictor}, we obtain
\[
\begin{aligned}
\boldsymbol{\Sigma}_{rZZ}
&=E(r\mathbf{Z}\mathbf{Z}^\top)\\
&=E\!\left[
r\boldsymbol{\Sigma}^{-1/2}
(\mathbf{X}-\boldsymbol{\mu})
(\mathbf{X}-\boldsymbol{\mu})^\top
\boldsymbol{\Sigma}^{-1/2}
\right]\\
&=E\!\left[
\boldsymbol{\Sigma}^{-1/2}r
(\mathbf{X}-\boldsymbol{\mu})
(\mathbf{X}-\boldsymbol{\mu})^\top
\boldsymbol{\Sigma}^{-1/2}
\right]
\quad\text{(since $r$ is scalar)}\\
&=\boldsymbol{\Sigma}^{-1/2}
E[r(\mathbf{X}-\boldsymbol{\mu})
(\mathbf{X}-\boldsymbol{\mu})^\top]
\boldsymbol{\Sigma}^{-1/2}\\
&=\boldsymbol{\Sigma}^{-1/2}\mathbf{T}_r
\boldsymbol{\Sigma}^{-1/2}.
\end{aligned}
\]
Thus,
\[
\begin{aligned}
\mathbf{M}_{\mathrm{PHD},r}
&=\boldsymbol{\Sigma}^{1/2}
\boldsymbol{\Sigma}_{rZZ}^{2}\boldsymbol{\Sigma}^{1/2}\\
&=\boldsymbol{\Sigma}^{1/2}
\bigl\{\boldsymbol{\Sigma}^{-1/2}\mathbf{T}_r
\boldsymbol{\Sigma}^{-1}\mathbf{T}_r
\boldsymbol{\Sigma}^{-1/2}\bigr\}
\boldsymbol{\Sigma}^{1/2}\\
&=\mathbf{T}_r\boldsymbol{\Sigma}^{-1}\mathbf{T}_r.
\end{aligned}
\]

To express $\mathbf{T}_r$ in terms of original-coordinate moments,
we write $r=Y-E(Y)-\{E(Y\mathbf{Z})\}^\top\mathbf{Z}$.
Using \eqref{eq:standardized-predictor}, we first express
$E(Y\mathbf{Z})$ as
\[
\begin{aligned}
E(Y\mathbf{Z})
&=E\{Y\boldsymbol{\Sigma}^{-1/2}
(\mathbf{X}-\boldsymbol{\mu})\}\\
&=E\{\boldsymbol{\Sigma}^{-1/2}
Y(\mathbf{X}-\boldsymbol{\mu})\}
\quad\text{(since $Y$ is scalar)}\\
&=\boldsymbol{\Sigma}^{-1/2}
E\{Y(\mathbf{X}-\boldsymbol{\mu})\}\\
&=\boldsymbol{\Sigma}^{-1/2}
\{E(\mathbf{X}Y)-\boldsymbol{\mu}E(Y)\}\\
&=\boldsymbol{\Sigma}^{-1/2}
\operatorname{Cov}(\mathbf{X},Y).
\end{aligned}
\]
Thus,
\[
\begin{aligned}
\{E(Y\mathbf{Z})\}^\top\mathbf{Z}
&=\operatorname{Cov}(\mathbf{X},Y)^\top
\boldsymbol{\Sigma}^{-1/2}
\boldsymbol{\Sigma}^{-1/2}(\mathbf{X}-\boldsymbol{\mu})\\
&=\operatorname{Cov}(\mathbf{X},Y)^\top
\boldsymbol{\Sigma}^{-1}(\mathbf{X}-\boldsymbol{\mu})\\
&=\mathbf{b}^\top(\mathbf{X}-\boldsymbol{\mu}),
\end{aligned}
\]
where we define $\mathbf{b}=\boldsymbol{\Sigma}^{-1}
\operatorname{Cov}(\mathbf{X},Y)$.
Substituting this expression into the definition of $\mathbf{T}_r$
gives
\[
\begin{aligned}
\mathbf{T}_r
&=E\!\left[
\{Y-E(Y)-\mathbf{b}^\top(\mathbf{X}-\boldsymbol{\mu})\}
(\mathbf{X}-\boldsymbol{\mu})
(\mathbf{X}-\boldsymbol{\mu})^\top
\right]\\
&=E[\{Y-E(Y)\}(\mathbf{X}-\boldsymbol{\mu})
(\mathbf{X}-\boldsymbol{\mu})^\top]\\
&\quad-E\!\left[
\{\mathbf{b}^\top(\mathbf{X}-\boldsymbol{\mu})\}
(\mathbf{X}-\boldsymbol{\mu})
(\mathbf{X}-\boldsymbol{\mu})^\top
\right]\\
&=\mathbf{T}_Y
-E\!\left[
\{\mathbf{b}^\top(\mathbf{X}-\boldsymbol{\mu})\}
(\mathbf{X}-\boldsymbol{\mu})
(\mathbf{X}-\boldsymbol{\mu})^\top
\right].
\end{aligned}
\]
Since $\mathbf{b}$ is fixed and
$\mathbf{b}^\top(\mathbf{X}-\boldsymbol{\mu})
=\sum_{\ell=1}^{p}b_\ell(X_\ell-\mu_\ell)$,
\[
\begin{aligned}
\mathbf{T}_r
&=\mathbf{T}_Y
-\sum_{\ell=1}^{p}b_\ell
E[(X_\ell-\mu_\ell)(\mathbf{X}-\boldsymbol{\mu})
(\mathbf{X}-\boldsymbol{\mu})^\top]\\
&=\mathbf{T}_Y-\sum_{\ell=1}^{p}b_\ell\mathbf{C}_\ell,
\end{aligned}
\]
where
$\mathbf{C}_\ell
=E[(X_\ell-\mu_\ell)(\mathbf{X}-\boldsymbol{\mu})
(\mathbf{X}-\boldsymbol{\mu})^\top]$.

\newpage

\subsection{Method-Specific Kernel Estimation}
\label{app:available-case-moments}

\paragraph{SIR and SAVE.}
The population kernels derived in Section~\ref{app:population-kernels} are $\mathbf{M}_{\mathrm{SIR}}=\operatorname{Cov}\{E(\mathbf{X}\mid Y)\}$ and $\mathbf{M}_{\mathrm{SAVE}}=E[\{\boldsymbol{\Sigma}-\boldsymbol{\Gamma}(Y)\}\boldsymbol{\Sigma}^{-1}\{\boldsymbol{\Sigma}-\boldsymbol{\Gamma}(Y)\}]$, where $\boldsymbol{\Gamma}(Y)=\operatorname{Cov}(\mathbf{X}\mid Y)$. 

Following the slicing approach of SIR and SAVE, we divide the observations into $H$ approximately equal-sized slices according to the response values. Let $\mathcal{I}_h$ contain the subject indices in slice $h$, and let $\widehat p_h=|\mathcal{I}_h|/n$, where $n$ is the number of subjects used for fitting. Define
\[
\mathcal{S}_{h,j}=\mathcal{I}_h\cap\mathcal{S}_j,
\qquad
\mathcal{S}_{h,jt}=\mathcal{I}_h\cap\mathcal{S}_{jt},
\]
with $n_{h,j}=|\mathcal{S}_{h,j}|$ and $n_{h,jt}=|\mathcal{S}_{h,jt}|$.

For SIR, we approximate $\operatorname{Cov}\{E(\mathbf{X}\mid Y)\}$
by the weighted outer products of centered slice means. We estimate the
overall mean using all available observations for each predictor:
\[
\widetilde\mu_j
=\frac{1}{n_j}\sum_{i\in\mathcal{S}_j}x_{ij},
\qquad
\widetilde{\boldsymbol{\mu}}
=[\widetilde\mu_1,\ldots,\widetilde\mu_p]^\top.
\]
Similarly, we estimate the mean within slice $h$ using all available observations for each predictor in that slice:
\[
\widetilde\mu_{h,j}
=\frac{1}{n_{h,j}}\sum_{i\in\mathcal{S}_{h,j}}x_{ij},
\qquad
\widetilde{\boldsymbol{\mu}}_{(h)}
=[\widetilde\mu_{h,1},\ldots,\widetilde\mu_{h,p}]^\top.
\]
Substituting these estimates gives
\[
\widehat{\mathbf{M}}_{\mathrm{SIR}}
=\sum_{h=1}^{H}\widehat p_h
(\widetilde{\boldsymbol{\mu}}_{(h)}
-\widetilde{\boldsymbol{\mu}})
(\widetilde{\boldsymbol{\mu}}_{(h)}
-\widetilde{\boldsymbol{\mu}})^\top.
\]

For SAVE, we approximate $\boldsymbol{\Gamma}(Y)$ by the covariance within each slice. Define the means on the joint observation set by
\[
\widetilde\mu_{h,j}^{(jt)}
=\frac{1}{n_{h,jt}}
\sum_{i\in\mathcal{S}_{h,jt}}x_{ij},
\qquad
\widetilde\mu_{h,t}^{(jt)}
=\frac{1}{n_{h,jt}}
\sum_{i\in\mathcal{S}_{h,jt}}x_{it}.
\]
We estimate each covariance entry using all jointly observed values of the corresponding predictor pair within the slice:
\[
(\widetilde{\boldsymbol{\Gamma}}_{(h)})_{jt}
=\frac{1}{n_{h,jt}-1}
\sum_{i\in\mathcal{S}_{h,jt}}
(x_{ij}-\widetilde\mu_{h,j}^{(jt)})
(x_{it}-\widetilde\mu_{h,t}^{(jt)}).
\]
Thus, each covariance entry is centered using means from the same joint observation set. Replacing $\boldsymbol{\Sigma}$ with $\widehat{\boldsymbol{\Sigma}}$ and averaging over slices gives
\[
\widehat{\mathbf{M}}_{\mathrm{SAVE}}
=\sum_{h=1}^{H}\widehat p_h
(\widehat{\boldsymbol{\Sigma}}
-\widetilde{\boldsymbol{\Gamma}}_{(h)})
\widehat{\boldsymbol{\Sigma}}^{-1}
(\widehat{\boldsymbol{\Sigma}}
-\widetilde{\boldsymbol{\Gamma}}_{(h)}).
\]

\paragraph{$\boldsymbol{y}$-based PHD.}
The population kernel derived in Section~\ref{app:population-kernels} is $\mathbf{M}_{\mathrm{PHD},y}=\mathbf{T}_Y\boldsymbol{\Sigma}^{-1}\mathbf{T}_Y$, where $\mathbf{T}_Y=E[\{Y-E(Y)\}(\mathbf{X}-\boldsymbol{\mu})(\mathbf{X}-\boldsymbol{\mu})^\top]$. Its $(j,t)$ entry is the centered third moment
\[
(\mathbf{T}_Y)_{jt}
=E[\{Y-E(Y)\}(X_j-\mu_j)(X_t-\mu_t)].
\]

Because $Y$ is fully observed, we estimate this entry using
$\mathcal{S}_{jt}$, the observations for which both $X_j$ and
$X_t$ are observed. Define the means on this set by
\[
\widetilde\mu_Y^{(jt)}
=\frac{1}{n_{jt}}\sum_{i\in\mathcal{S}_{jt}}y_i,
\qquad
\widetilde\mu_j^{(jt)}
=\frac{1}{n_{jt}}\sum_{i\in\mathcal{S}_{jt}}x_{ij},
\qquad
\widetilde\mu_t^{(jt)}
=\frac{1}{n_{jt}}\sum_{i\in\mathcal{S}_{jt}}x_{it}.
\]
The available-case moment estimator is
\[
(\widetilde{\mathbf{T}}_Y)_{jt}
=\frac{n_{jt}}{(n_{jt}-1)(n_{jt}-2)}
\sum_{i\in\mathcal{S}_{jt}}
(y_i-\widetilde\mu_Y^{(jt)})
(x_{ij}-\widetilde\mu_j^{(jt)})
(x_{it}-\widetilde\mu_t^{(jt)}).
\]

The coefficient corrects the finite-sample bias from sample centering.
Replacing $\mathbf{T}_Y$ with $\widetilde{\mathbf{T}}_Y$ and
$\boldsymbol{\Sigma}$ with $\widehat{\boldsymbol{\Sigma}}$ gives
\[
\widehat{\mathbf{M}}_{\mathrm{PHD},y}
=\widetilde{\mathbf{T}}_Y
\widehat{\boldsymbol{\Sigma}}^{-1}
\widetilde{\mathbf{T}}_Y.
\]

\paragraph{$\boldsymbol{r}$-based PHD.}
The population kernel derived in Section~\ref{app:population-kernels} is $\mathbf{M}_{\mathrm{PHD},r}=\mathbf{T}_r\boldsymbol{\Sigma}^{-1}\mathbf{T}_r$, where $\mathbf{T}_r=\mathbf{T}_Y-\sum_{\ell=1}^{p}b_\ell\mathbf{C}_\ell$, $\mathbf{b}=\boldsymbol{\Sigma}^{-1}\operatorname{Cov}(\mathbf{X},Y)$, and $\mathbf{C}_\ell=E[(X_\ell-\mu_\ell)(\mathbf{X}-\boldsymbol{\mu})(\mathbf{X}-\boldsymbol{\mu})^\top]$. We estimate $\mathbf{b}$ and $\mathbf{C}_\ell$ separately and combine them with $\widetilde{\mathbf{T}}_Y$.

First, we estimate the predictor--response covariance vector using all available observations for each predictor. Since $Y$ is fully observed, the observation set for $\operatorname{Cov}(X_j,Y)$ is $\mathcal{S}_j$. Define the response mean on this set by
\[
\widetilde\mu_Y^{(j)}
=\frac{1}{n_j}\sum_{i\in\mathcal{S}_j}y_i.
\]
Using the predictor mean $\widetilde\mu_j$ defined above, we obtain
\[
\widetilde c_j
=\frac{1}{n_j-1}\sum_{i\in\mathcal{S}_j}
(x_{ij}-\widetilde\mu_j)(y_i-\widetilde\mu_Y^{(j)}).
\]
Collecting these estimates gives
$\widetilde{\mathbf{c}}=[\widetilde c_1,\ldots,\widetilde c_p]^\top$.
Replacing $\boldsymbol{\Sigma}$ with $\widehat{\boldsymbol{\Sigma}}$, we estimate the regression coefficient by $\widehat{\mathbf{b}}=\widehat{\boldsymbol{\Sigma}}^{-1}\widetilde{\mathbf{c}}$.

Next, the $(j,t)$ entry of $\mathbf{C}_\ell$ is the centered third moment
\[
(\mathbf{C}_\ell)_{jt}
=E[(X_\ell-\mu_\ell)(X_j-\mu_j)(X_t-\mu_t)].
\]
We estimate this entry using all observations for which $X_j$, $X_t$, and $X_\ell$ are jointly observed. Define
\[
\mathcal{S}_{jt\ell}
=\mathcal{S}_j\cap\mathcal{S}_t\cap\mathcal{S}_\ell,
\qquad
n_{jt\ell}=|\mathcal{S}_{jt\ell}|.
\]
The means on this joint observation set are
\[
\widetilde\mu_j^{(jt\ell)}
=\frac{1}{n_{jt\ell}}
\sum_{i\in\mathcal{S}_{jt\ell}}x_{ij},
\qquad
\widetilde\mu_t^{(jt\ell)}
=\frac{1}{n_{jt\ell}}
\sum_{i\in\mathcal{S}_{jt\ell}}x_{it},
\qquad
\widetilde\mu_\ell^{(jt\ell)}
=\frac{1}{n_{jt\ell}}
\sum_{i\in\mathcal{S}_{jt\ell}}x_{i\ell}.
\]
The available-case moment estimator is
\[
(\widetilde{\mathbf{C}}_\ell)_{jt}
=\frac{n_{jt\ell}}{(n_{jt\ell}-1)(n_{jt\ell}-2)}
\sum_{i\in\mathcal{S}_{jt\ell}}
(x_{ij}-\widetilde\mu_j^{(jt\ell)})
(x_{it}-\widetilde\mu_t^{(jt\ell)})
(x_{i\ell}-\widetilde\mu_\ell^{(jt\ell)}).
\]

The coefficient applies the same finite-sample correction
as in $y$-based PHD. Substituting these estimates into the
residual moment decomposition gives
\[
\widetilde{\mathbf{T}}_r
=\widetilde{\mathbf{T}}_Y
-\sum_{\ell=1}^{p}\widehat b_\ell\widetilde{\mathbf{C}}_\ell.
\]
Replacing $\mathbf{T}_r$ with $\widetilde{\mathbf{T}}_r$ and
$\boldsymbol{\Sigma}$ with $\widehat{\boldsymbol{\Sigma}}$ then gives
\[
\widehat{\mathbf{M}}_{\mathrm{PHD},r}
=\widetilde{\mathbf{T}}_r
\widehat{\boldsymbol{\Sigma}}^{-1}
\widetilde{\mathbf{T}}_r.
\]

\newpage

\section{Sufficient Subspaces and Prediction Risk}
\label{app:reduced-risk}

Suppose $E(Y^2)<\infty$. For a loading matrix $\mathbf{A}$,
define the optimal prediction risk using the subspace it spans:
\[
\mathcal{R}(\mathbf{A})
=\inf_g E[\{Y-g(\mathbf{A}^\top\mathbf{X})\}^2],
\]
where the infimum is over all measurable functions $g$.

\paragraph{Proposition.}
If $\operatorname{col}(\mathbf{B}_0)$ is a sufficient subspace, then
\[
\mathcal{R}(\mathbf{B}_0)
=E\{\operatorname{Var}(Y\mid\mathbf{X})\}
\leq \mathcal{R}(\mathbf{A})
\]
for every candidate loading matrix $\mathbf{A}$. Thus, a sufficient subspace attains the minimum
prediction risk under squared-error loss when evaluated
using its optimal predictor.

\begin{proof}
For any candidate loading matrix $\mathbf{A}$ and predictor $g$, the law of total expectation gives
\[
E[\{Y-g(\mathbf{A}^\top\mathbf{X})\}^2]
=E\!\left[E[\{Y-g(\mathbf{A}^\top\mathbf{X})\}^2\mid\mathbf{X}]\right].
\]
Let $m(\mathbf{X})=E(Y\mid\mathbf{X})$. Write $Y-g(\mathbf{A}^\top\mathbf{X})=\{Y-m(\mathbf{X})\}+\{m(\mathbf{X})-g(\mathbf{A}^\top\mathbf{X})\}$ and note that the second term is fixed conditional on $\mathbf{X}$. Squaring and taking conditional expectations, we obtain
\[
\begin{aligned}
E[\{Y-g(\mathbf{A}^\top\mathbf{X})\}^2\mid\mathbf{X}]
&=E[\{Y-m(\mathbf{X})\}^2\mid\mathbf{X}]\\
&\quad+2\{m(\mathbf{X})-g(\mathbf{A}^\top\mathbf{X})\}
          E\{Y-m(\mathbf{X})\mid\mathbf{X}\}\\
&\quad+\{m(\mathbf{X})-g(\mathbf{A}^\top\mathbf{X})\}^2.
\end{aligned}
\]
Since $E\{Y-m(\mathbf{X})\mid\mathbf{X}\}=0$, the cross term vanishes. Consequently,
\[
E[\{Y-g(\mathbf{A}^\top\mathbf{X})\}^2\mid\mathbf{X}]
=\operatorname{Var}(Y\mid\mathbf{X})
+\{m(\mathbf{X})-g(\mathbf{A}^\top\mathbf{X})\}^2.
\]
Taking outer expectations gives
\begin{equation}
\label{eq:population-risk}
\begin{aligned}
E[\{Y-g(\mathbf{A}^\top\mathbf{X})\}^2]
&=E\{\operatorname{Var}(Y\mid\mathbf{X})\}\\
&\quad+E[\{m(\mathbf{X})-g(\mathbf{A}^\top\mathbf{X})\}^2]\\
&\geq E\{\operatorname{Var}(Y\mid\mathbf{X})\}.
\end{aligned}
\end{equation}
Taking the infimum over $g$ therefore yields $\mathcal{R}(\mathbf{A})\geq E\{\operatorname{Var}(Y\mid\mathbf{X})\}$.

Now suppose $\operatorname{col}(\mathbf{B}_0)$ is sufficient. Since $Y\perp\mathbf{X}\mid\mathbf{B}_0^\top\mathbf{X}$ and $\mathbf{B}_0^\top\mathbf{X}$ is a function of $\mathbf{X}$, we have $E(Y\mid\mathbf{B}_0^\top\mathbf{X})=E(Y\mid\mathbf{X},\mathbf{B}_0^\top\mathbf{X})=E(Y\mid\mathbf{X})=m(\mathbf{X})$ almost surely. Choosing this conditional mean as the predictor makes the second term in \eqref{eq:population-risk} zero, so $\mathcal{R}(\mathbf{B}_0)=E\{\operatorname{Var}(Y\mid\mathbf{X})\}$. Thus, a sufficient subspace attains the lower bound. More generally, the central mean subspace also attains this lower bound because it preserves the conditional mean.

However, the converse does not hold in general. Let $X_1,X_2,\epsilon$ be independent standard normal variables and set $\mathbf{X}=(X_1,X_2)^\top$ and $Y=X_1+X_2\epsilon$. Using only $X_1$ attains the minimum risk because $E(Y\mid\mathbf{X})=E(Y\mid X_1)=X_1$. However, $\operatorname{Var}(Y\mid\mathbf{X})=X_2^2$, whereas $\operatorname{Var}(Y\mid X_1)=\operatorname{Var}(X_2\epsilon\mid X_1)=\operatorname{Var}(X_2\epsilon)=1$. If $\operatorname{span}\{(1,0)^\top\}$ were sufficient, the conditional distributions of $Y$ given $\mathbf{X}$ and given $X_1$ would coincide, and hence their variances would agree almost surely. Since $X_2^2\neq1$ almost surely, this subspace is not sufficient. Thus, $\operatorname{span}\{(1,0)^\top\}$ is the central mean subspace but is not a sufficient subspace.
\end{proof}

\newpage

\section{Distance Invariance through Orthonormalization}
\label{app:basis-invariance}

\begin{lemma}
\label{lem:basis-invariance}
Let $\mathbf{Q}_1,\mathbf{Q}_2\in\mathbb{R}^{p\times d}$
be orthonormal bases of the same subspace. For any
$\mathbf{x}_i,\mathbf{x}_j\in\mathbb{R}^p$,
\[
\|\mathbf{Q}_1^\top(\mathbf{x}_i-\mathbf{x}_j)\|_2
=\|\mathbf{Q}_2^\top(\mathbf{x}_i-\mathbf{x}_j)\|_2.
\]
\end{lemma}
\begin{proof}
Since the bases span the same subspace,
$\mathbf{Q}_2=\mathbf{Q}_1\mathbf{R}$ for some
$\mathbf{R}\in\mathbb{R}^{d\times d}$. Thus,
\[
\mathbf{Q}_1^\top\mathbf{Q}_2
=\mathbf{Q}_1^\top\mathbf{Q}_1\mathbf{R}
=\mathbf{R}.
\]
Note that $\mathbf{Q}_1\mathbf{Q}_1^\top$ is the orthogonal
projection matrix onto $\operatorname{col}(\mathbf{Q}_1)$, since
\(
\mathbf{Q}_1\mathbf{Q}_1^\top
=\mathbf{Q}_1(\mathbf{Q}_1^\top\mathbf{Q}_1)^{-1}\mathbf{Q}_1^\top.
\)
The columns of $\mathbf{Q}_2$ lie in the same subspace, so
$\mathbf{Q}_1\mathbf{Q}_1^\top\mathbf{Q}_2=\mathbf{Q}_2$.
Thus,
\[
\mathbf{R}^\top\mathbf{R}
=\mathbf{Q}_2^\top\mathbf{Q}_1\mathbf{Q}_1^\top\mathbf{Q}_2
=\mathbf{Q}_2^\top\mathbf{Q}_2
=\mathbf{I}_d.
\]
Therefore, $\mathbf{R}$ is orthogonal and its transpose
preserves Euclidean norms, so
\[
\|\mathbf{Q}_2^\top(\mathbf{x}_i-\mathbf{x}_j)\|_2
=\|\mathbf{R}^\top\mathbf{Q}_1^\top(\mathbf{x}_i-\mathbf{x}_j)\|_2
=\|\mathbf{Q}_1^\top(\mathbf{x}_i-\mathbf{x}_j)\|_2.
\]
\end{proof}

\newpage

\section{Additional Simulation Results}
\label{app:additional-simulations}

The three additional simulations follow the design in
Section~\ref{sec:simulation-design}, using the following
response models for SIR, SAVE, and $r$-based PHD:
\[
\begin{aligned}
\text{SIR:}\quad
&Y=\eta_1(\eta_1+\eta_2+1)+0.2\epsilon,\\
\text{SAVE:}\quad
&Y=\cos(2\eta_1)-\cos(\eta_2)+0.2\epsilon,\\
\text{$r$-based PHD:}\quad
&Y=\eta_1+\eta_2+2\eta_1^2+\eta_2^2+0.2\epsilon.
\end{aligned}
\]
Table~\ref{tab:additional-results} reports the projection
distances. The proposed estimator achieves the lowest mean
projection distance in all 27 additional settings.

\input{tb_simulation_additional}

%% file: tb_simulation_additional.tex
\begin{table}[H]
\centering
\setstretch{1}
\footnotesize
\setlength{\tabcolsep}{3pt}
\renewcommand{\arraystretch}{1.0}
\begin{tabular*}{\textwidth}{@{\extracolsep{\fill}}llrrrr@{}}
\toprule
Cov. & Pattern & Complete & Available & SoftImpute & \shortstack{Shrinkage\\(proposed)} \\
\midrule
\multicolumn{6}{@{}l}{\textit{SIR: quadratic interaction}} \\
\addlinespace[2pt]
AR1 & P1 & 1.594 (0.186) & 1.503 (0.206) & 1.401 (0.197) & \textbf{1.047} (0.276) \\
AR1 & P2 & 1.594 (0.186) & 1.456 (0.198) & 1.371 (0.216) & \textbf{0.960} (0.259) \\
AR1 & P3 & 1.594 (0.186) & 1.443 (0.211) & 1.324 (0.230) & \textbf{0.953} (0.286) \\
C1 & P1 & 1.644 (0.184) & 1.582 (0.194) & 1.495 (0.196) & \textbf{1.095} (0.324) \\
C1 & P2 & 1.644 (0.184) & 1.594 (0.177) & 1.470 (0.203) & \textbf{1.058} (0.321) \\
C1 & P3 & 1.644 (0.184) & 1.544 (0.196) & 1.428 (0.211) & \textbf{1.018} (0.321) \\
C2 & P1 & 1.700 (0.160) & 1.590 (0.138) & 1.578 (0.163) & \textbf{1.136} (0.280) \\
C2 & P2 & 1.700 (0.160) & 1.653 (0.125) & 1.540 (0.177) & \textbf{1.097} (0.278) \\
C2 & P3 & 1.700 (0.160) & 1.644 (0.161) & 1.544 (0.187) & \textbf{1.071} (0.269) \\
\midrule
\multicolumn{6}{@{}l}{\textit{SAVE: cosine difference}} \\
\addlinespace[2pt]
AR1 & P1 & 1.683 (0.107) & 1.704 (0.099) & 1.651 (0.121) & \textbf{1.508} (0.226) \\
AR1 & P2 & 1.683 (0.107) & 1.684 (0.095) & 1.642 (0.094) & \textbf{1.531} (0.168) \\
AR1 & P3 & 1.683 (0.107) & 1.672 (0.090) & 1.643 (0.101) & \textbf{1.570} (0.079) \\
C1 & P1 & 1.717 (0.105) & 1.757 (0.110) & 1.635 (0.097) & \textbf{1.453} (0.201) \\
C1 & P2 & 1.717 (0.105) & 1.792 (0.118) & 1.662 (0.093) & \textbf{1.492} (0.128) \\
C1 & P3 & 1.717 (0.105) & 1.766 (0.095) & 1.653 (0.096) & \textbf{1.501} (0.096) \\
C2 & P1 & 1.694 (0.124) & 1.782 (0.138) & 1.610 (0.108) & \textbf{1.358} (0.206) \\
C2 & P2 & 1.694 (0.124) & 1.805 (0.138) & 1.618 (0.090) & \textbf{1.363} (0.200) \\
C2 & P3 & 1.694 (0.124) & 1.788 (0.130) & 1.623 (0.098) & \textbf{1.418} (0.180) \\
\midrule
\multicolumn{6}{@{}l}{\textit{$r$-based PHD: linear-plus-quadratic}} \\
\addlinespace[2pt]
AR1 & P1 & 1.458 (0.149) & 1.411 (0.172) & 1.319 (0.182) & \textbf{0.971} (0.207) \\
AR1 & P2 & 1.458 (0.149) & 1.364 (0.176) & 1.276 (0.194) & \textbf{0.937} (0.198) \\
AR1 & P3 & 1.458 (0.149) & 1.327 (0.173) & 1.263 (0.188) & \textbf{0.914} (0.200) \\
C1 & P1 & 1.467 (0.159) & 1.526 (0.150) & 1.245 (0.183) & \textbf{0.955} (0.192) \\
C1 & P2 & 1.467 (0.159) & 1.488 (0.204) & 1.199 (0.189) & \textbf{0.909} (0.196) \\
C1 & P3 & 1.467 (0.159) & 1.450 (0.181) & 1.196 (0.181) & \textbf{0.851} (0.167) \\
C2 & P1 & 1.651 (0.110) & 1.760 (0.087) & 1.477 (0.144) & \textbf{1.142} (0.204) \\
C2 & P2 & 1.651 (0.110) & 1.708 (0.097) & 1.457 (0.154) & \textbf{1.092} (0.199) \\
C2 & P3 & 1.651 (0.110) & 1.689 (0.116) & 1.429 (0.161) & \textbf{1.094} (0.212) \\
\bottomrule
\end{tabular*}
\caption{Projection distances (mean and SD) over successful fits
for three additional response models. Bold indicates the lowest
mean in each setting.}
\label{tab:additional-results}
\end{table}